\documentclass{aastex61}
\usepackage{enumitem,amsmath}
\usepackage{dutchcal}

\def\W01{Will_astroph_2001}

\newcommand{\half}{\frac{1}{2}}
\newcommand{\myem}{\it}
\newcommand{\del}{{\bf \nabla}}
\newcommand{\beq}{\begin{equation}}
\newcommand{\eeq}{\end{equation}}

\newcommand{\vR}{v_{{}\!_R}}
\newcommand{\SD}{\Sigma_{d}}
\newcommand{\CL}{C_L}
\newcommand{\Mdd}{\dot M_d}
\newcommand{\Mdj}{\dot M_j}
\newcommand{\Ldd}{\dot L_d}
\newcommand{\Ldj}{\dot L_j}
\newcommand{\Lacc}{{\cal L}_{\rm acc}}

\newcommand{\cool}{\color{iceberg}}

\newcommand{\simge}{\gtrsim}
\newcommand{\simle}{\lesssim}

\newcommand{\sie}{e} 

\newcommand{\Wtot}{\mathbb{W}}

\newcommand{\Ptot}{\mathbb{P}}
\newcommand{\Rtot}{\mathbb{R}}
\newcommand{\LO}{{\cal L}_0}
\newcommand{\LP}{{\cal L}'_0}

\definecolor{gray}{rgb}{0.5, 0.5, 0.5}
\definecolor{iceberg}{rgb}{0.44, 0.65, 0.82}
\definecolor{mygreen}{rgb}{0.05, 0.55, 0.05}

\begin{document}

\title{Turbulent accretion braking torques and efficient jets without magnetocentrifugal acceleration: Core concepts.}
\author{Peter Todd Williams}


\begin{abstract}
I discuss three mutually-supportive notions or assumptions regarding jets and accretion. The first is magnetocentrifugal acceleration (MCA), the overwhelmingly favored mechanism for the production of jets in most steady
accreting systems. The second is the zero-torque inner boundary condition. The third is that effective viscous dissipation is like real dissipation, leading directly to heating. All three assumptions fit nicely together in a manner that is simple, persuasive, and mutually-consistent. All, I argue, are incorrect. For concreteness I focus on protostars.
Magnetohydrodynamic (MHD) turbulence in accretion is not a sink of energy, but a reservoir, capable of doing mechanical work directly and therefore efficiently, rather than solely through ohmic (``viscous'') heating. Advection of turbulence energy reduces the effective radiative efficiency, and may help solve the missing boundary-layer emission problem. The angular momentum problem, whereby accretion spins up a protostar to breakup, is resolved by allowing direct viscous coupling to the protostar, permitting substantially greater energy to be deposited into the accretion flow than otherwise possible. This goes not into heat, but into a turbulent, tangled, buoyant toroidal magnetic field.
I argue that there is neither an angular momentum problem nor an efficiency problem that MCA is needed to solve. Moreover, the turbulent magnetic field has ample strength not just to collimate but to accelerate gas, first radially inwards through tension forces and then vertically through pressure forces, without any MCA mechanism. I suggest then that jets,
particularly the most powerful and well-collimated protostellar jets, are not magnetocentrifugally driven.
\end{abstract}

\section{Introduction}
For many years now, the conventional wisdom regarding most astrophysical jets, especially persistent jets such as in protostars, AGN and microquasars, is that they are driven by some type of magnetocentrifugal acceleration (MCA) by large-scale poloidal magnetic fields \citep{BP_1982}. Here however, as I have done consistently elsewhere, I take a contrarian view regarding the MCA hypothesis, and I continue the argument against it on theoretical grounds. I focus on the physics of the innermost accretion regions in steady disk-mediated accretion onto a central object, particularly, a protostar. 

Conventionally, accretion onto a protostar is thought to occur in one of two scenarios. Either the accretion occurs in a radially-thin boundary layer that does not viscously transmit torque between the protostar and the surrounding disk, or accretion is funneled from a truncated accretion disk onto the protostar by the protostar's magnetosphere \citep[see {\it e.g.} fig.~1 of][]{HerDubHur_2005}.

In the first scenario, accretion in a boundary layer leads to the so-called angular momentum problem, because the boundary layer can not transmit torque and the protostar gains excess angular momentum from pure advection. This halts accretion. Even before then, this scenario also predicts a larger UV or X-ray flux from the boundary layer than is typically seen. Magnetocentrifugal acceleration --- either from the protostar \citep{HarMGr_ApJ_1982} or from the surrounding nearby accretion flow, possibly including the broader disk as well \citep{PudNor_ApJ_1983} --- is an appealing solution to these problems, as well as the problem of jet production. The magnetocentrifugal mechanism provides a sink of angular momentum and thereby enables continued accretion of mass onto the protostar, while curbing the attendant accumulation of angular momentum that would otherwise tear the protostar apart. On the other hand it raises new problems, such as the origin of the large-scale poloidal field, and the mechanism by which matter is threaded onto it.

In the second scenario, material is funneled onto the protostar from more radially-distant regions of an accretion disk by a strong large-scale poloidal field anchored in the protostar. See \citet{Kon_ApJL_1991}, who extends the neutron-star accretion model of \citet{GhoLam_ApJL_1978} \citep[see also][]{GhoLamPet_ApJ_1977} to protostars. In that case, it is assumed the disk has an inner truncation radius corresponding to the coupling region, and so there is no boundary-layer. 
This potentially helps solve the angular momentum problem, because while the material at larger radii has even greater specific angular momentum than material in a boundary-layer would, the poloidal field can transmit torque to the disk, keeping the protostar from gaining excess angular momentum. Jets may still be driven magnetocentrifugally, such as by the disk or by the X-wind process, in which jets originate in this coupling region. On the other hand, this MCA suffers from the problem again of being dependent on assumptions about the reconnection process by which material is threaded onto open field lines, and it also relies upon assumptions about the field topology. 
As well, the observed speeds of jets, being of order of a few times the escape velocity at the protostellar surface, is suggestive of a jet origin very near the protostar, which tends to argue against disk-winds in particular for the high-velocity (jet) component of outflows.

\begin{figure}
 \centering
 \includegraphics[width=100mm]{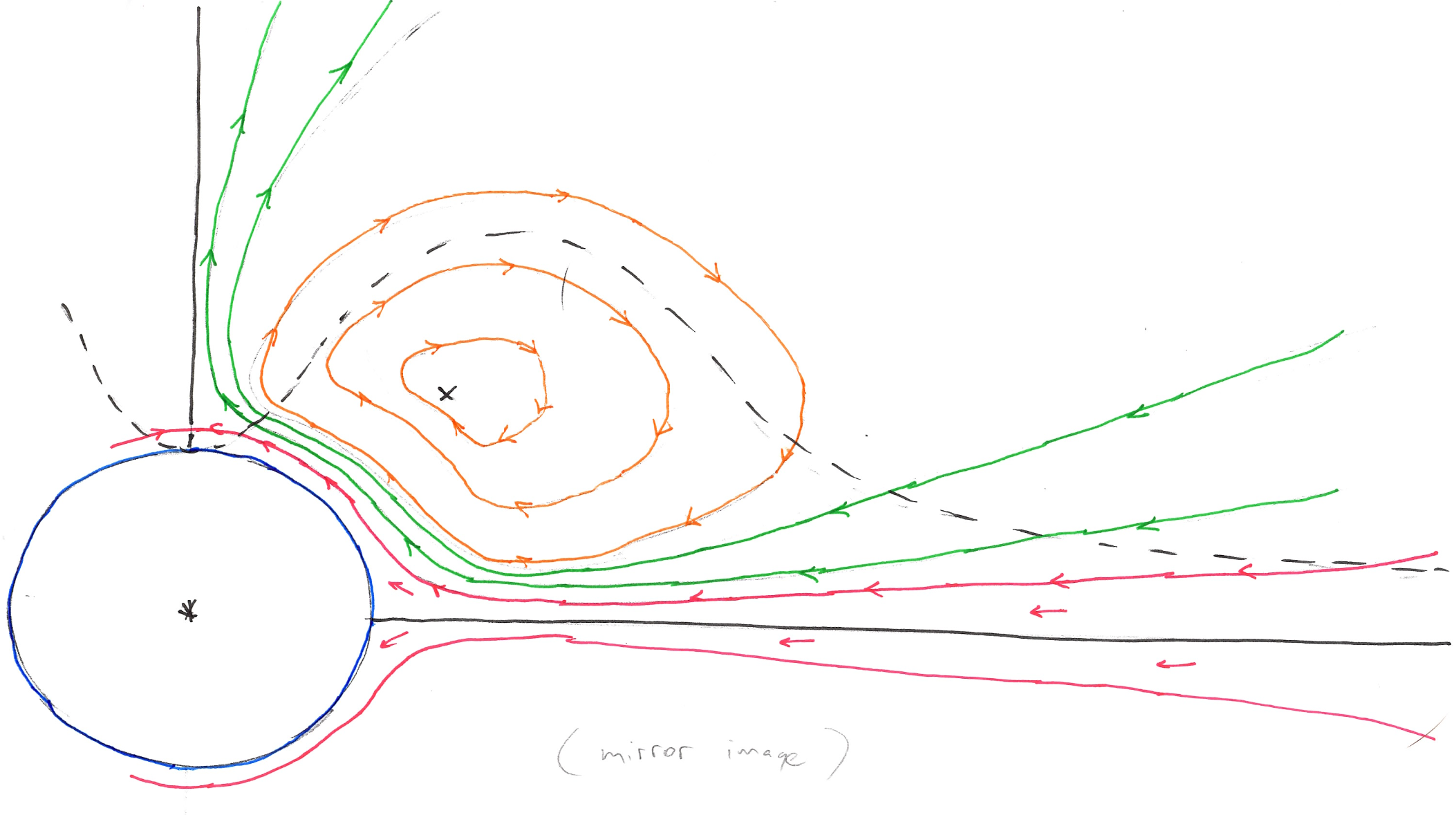}
  \caption{I hypothesize a protostar (blue) embedded in a surrounding geometrically- and optically-thick accretion flow or envelope. Streamlines are indicated with continuous arrowed lines. The dashed black line is representative of the effective photosphere of this system in or near optical/IR wavelengths. The envelope includes two large hot ($T \simle 10^4\ {\rm K}$) recirculation zones above and below the disk (orange). The red streamline is the separatrix; it separates that part of the flow that ultimately accretes onto the protostar (red arrows) from the flow that does not (green streamlines). Veiling is not due to accretion shocks, but the recirculation zones. (NOTE: {\it I am awaiting artwork provided by a professional illustrator for this and other line drawings in this submission.}) }
  \label{fig:2D_recirc}
\end{figure}

In both cases, some flavor of MCA has been suggested to help solve the angular momentum problem by providing a sink for angular momentum in the form of a jet. 
If, alternatively, the disk couples viscously to the protostar, then there is no angular momentum problem to be solved in the first place.
Indeed, \citet{Pri_1989} appears to make a similar argument in his objections to \citet{Shu_etal_1988}, the intellectual precursor to the magnetocentrifugal X-wind model of \citet{Shu_etal_ApJ_1994}.
Particularly in the case of young or high accretion-rate protostars that are spinning at a substantial fraction (e.g. $\gtrsim$ 30\%) of breakup, I therefore suggest a third scenario, in which accretion occurs through a geometrically thick flow that extends all the way down to the surface of the protostar and envelopes it, so that the protostar is essentially fully embedded in the flow (fig.\ \ref{fig:2D_recirc}).
(I will continue to use the terms ``star'' and ``disk,'' but with the very important caveat that now there may not actually be a clearly delineable distinction between the two.)
In this scenario there is no radially-thin boundary layer and no zero-torque boundary condition on the disk. The protostar is able to exchange torques with the disk viscously, through the tangled, turbulent magnetic field generated by the magnetorotational instability (MRI), possibly augmented by convective or other modes. This viscous coupling of the protostar to the disk allows the protostar to gain mass but not excess angular momentum. No MCA is needed, nor any large-scale poloidal field onto which matter must somehow be coupled. Of course, dynamo processes can be expected to generate some type of large-scale organized field, but the details of this field are not integral to the general schema at this stage and are not addressed here.

It has long been recognized that the inward advection of material in this innermost region in steady accretion onto a protostar is an appealing source for the energy needed to drive jets and related bipolar outflows or winds; for an early discussion, see \cite{Pri_1989}; also see \cite{Tor_1984} and \cite{Tor_1986}. The reason for this appeal is that a substantial fraction 
of all accretion energy can potentially be released there; this is a simple consequence of the virial theorem.

For example, in the case there actually is a thin boundary layer, in the standard thin disk theory of slow ($v_R \ll v_\phi$), cool ($H \ll R$), radiatively efficient ($q^+ \simeq q^- \gg q_{\rm adv})$ accretion with negligible vertical mass loss, 
the radiative luminosity of this boundary layer 
is\footnote{Early work in the field gave a slightly different and subtly incorrect result for this luminosity.}
\beq
{\cal L}^{({\rm rad})}_{\rm BL} = \frac{GM_*\dot M}{2R_*}\left(1 - \frac{\Omega_*}{\kappa_*}\right)^2.
\label{eq:BLlum}
\eeq
To the extent that the innermost region of any real disk or other accretion flow is not thin, cool, radiatively efficient, or is otherwise not ideal, this relation may not hold, but it still serves as a useful benchmark for the order of magnitude of energy available there. Boundary layer or not, there is still an enormous amount of energy that potentially may be released in the inner accretion region, of at least comparable order to the relation given above. 

The binding energy that must be lost by accreting matter as it falls down the gravitational well towards the star is only part of the energy budget. The second appealing and conventional source of energy to power protostellar jets is the reservoir of rotational energy in the angular momentum of the star, such as again in X-wind models \citep[][{\em et seq}.]{Shu_etal_ApJ_1994}, in which mass loading and torque create a feedback process that maintains the protostellar rotation at some fixed fraction of breakup throughout the accretion process.

I too relied upon torques on the star to power an outflow in my own initial work \cite{\W01} on this problem, but in contrast with the X-wind model, I suggested the torques were transmitted viscously. The power available in the form of mechanical luminosity from viscously-coupled braking torque\footnote{Note that I take care not to call this braking torque a ``spin-down'' torque. This is potentially an important distinction for a central object with a rather soft equation of state, such as a protostar, as opposed to a neutron star with its very stiff equation of state. Particularly for a protostar, a gain or loss of angular momentum may be simultaneous with a gain or loss of moment of inertia, and so for example what otherwise might have been called a ``spin-down'' torque can actually result in spinning a body up, such as if the object simultaneously shrinks in effective radius or loses mass, perhaps even due to the same physical effect as that responsible for the torque in the first place.} from a fast-rotating protostar is substantial. Generally and roughly, like the boundary-layer luminosity above, the mechanical power available from the braking torque in steady accretion is of order
\beq
{\cal L}^{({\rm mech})}_{\rm BT} \approx G M_* \dot M / R_*,
\label{eq:BTorder}
\eeq
as in fact was recognized some time ago \citep[see][p.~634]{LBP}, and may be much larger still than the notional boundary-layer luminosity above. It is a commonplace observation that as much as half of all accretion energy may be released in the innermost regions of accretion (such as the boundary layer); actually, it is potentially even far more than that, as much as three times as much or even more in fact in the limiting case of a maximally-rotating central object. (The explanation for this apparently paradoxical result is that the energy nominally ``released'' or dissipated in the accretion flow can exceed the accretion budget thanks to the additional energy delivered from the star via turbulent stresses.)
In either case, whether from the last stages of accretion onto the protostar or from the protostar itself, there is abundantly sufficient energy to power observed jets.

Energy is not the problem. Efficient coupling is, especially if the jet power comes from heat; the Second Law will always take its share. 
The power that goes into jets is a sizeable fraction of the total accretion power budget; this alone sets a floor on the thermodynamic efficiency of the engine that drives jets. Moreover, if jets were driven thermally, as in a pressure-driven flow through a de\,Laval nozzle, instead of mechanically such as in the MCA hypothesis, then this would require high temperatures and associated large X-ray fluxes that are not seen; this again argues all the more so for an efficient mechanical means of powering jets. The MCA hypothesis addresses this by proposing a mechanically-efficient mechanism that does not rely upon thermal driving.

So it might appear that a key problem in powering outflows from an alternative mechanism that relies upon either of these two sources appealed to above (the energy of accretion in the innermost regions or the rotation of the protostar itself) is that energy in either case is extracted ``viscously'' --- that is, through the action of the turbulence and its effective ``viscosity'' --- and therefore leads to heating. This is incorrect. As I pointed out in \cite{\W01} and later in \cite{Will_MNRAS_2005}, the energy that goes into the turbulence creates a tangled, toroidal magnetic field, and the associated hoop-stresses create a radially-inwards force that is able to do real work on the accreting material. In other words, the energy that goes into turbulence can come out in the form of mechanical work, not heat, not even as an intermediary.
Here, I take a step towards resolving both turbulence energetics and global energetics in further detail.

The assumption that turbulence always and instantaneously results in heating is present in both in standard thin-disk theory as well as in the theory of ADAFs, where it is common to write that there is a heating term corresponding to a local ``viscous dissipation'' rate (per area) of
\beq
{\cal Q}_+ = \nu_{\rm t} \Sigma R^2 (\Omega')^2 = \int_{-\infty}^{\infty} {\cal q}_+\ dz,
\label{eq:diss0}
\eeq
and where ${\cal q}_+$, the (per unit volume) viscous heating, ignoring compressibility for simplicity, may be written
\beq
{\cal q}_+ = \frac{1}{2\rho\nu_{\rm t}} W : W \simeq \rho \nu_{\rm t} W_{\langle R \phi \rangle}^2 = \rho \nu_{\rm t} R^2 \left( \Omega' \right)^2,
\label{eq:diss1}
\eeq
which is also\footnote{I use greek indices down and up to refer to components in the covariant/contravariant language of tensors, with a comma indicating a simple partial derivative and a semicolon indicating a covariant derivative. For cartesian index notation, I use general latin indices such as $i$, $j$, $k$, and I will use an explicit partial derivative symbol instead of a comma to indicate differentiation, e.g. $\partial_i u_j$ is shorthand for $\partial u_j / \partial x_i$. Angle brackets indicate an actual physical component, which can be related to covariant or contravariant quantities through the square root of the metric, as the cylindrical coordinate system is orthogonal, and no other coordinate system is used.}
\beq
{\cal q}_+ = v_{\alpha; \beta}W^{\alpha \beta} = \left( {\bf \nabla} {\bf v} \right) : W
\label{eq:diss2}
\eeq
where $W$ is the (turbulent) stress tensor\footnote{The thermodynamic pressure $P$ is not included stress tensor $W$ here; otherwise, in compressible flow, $(\nabla {\bf v}) : W$ also would include a $PdV$ work contribution, in addition to heating. In principle, $W$ does however include turbulent pressure, but in the interest of simplicity I will largely sidestep the issue of turbulent pressure as it detracts from the main points I am trying to assert in this paper.}, 
$\nu_{\rm t}$ is the effective turbulent viscosity, $\bf v$ is the velocity with covariant components $v_\alpha$, and the remaining symbols are standard.
See for example eq.~11.3 and also eq.~4.29 in \cite{FKR}, pages 609--610 in \cite{LBP} and page~344 of \cite{SS} in an accretion context, and eqs.~49.2-5 in \cite{LL6} for the fluid-dynamical basis of this in viscous fluids.

Again, this assumes that all of the energy that goes into turbulence comes out in the form of heat, and moreover, that it does so on a time scale (the dissipative time scale) that is short compared to the radial advective time scale.
Neither assumption is warranted. In other words, use of these relations for $q_+$ carries the analogy between real viscosity and effective turbulent viscosity too far. Because the distinction is so important, in this paper, I will always designate a turbulent effective kinematic viscosity as $\nu_{\rm t}$, and reserve $\nu$ exclusively for real molecular kinematic viscosity. The effective dissipation due to turbulence as represented in eqns.~\ref{eq:diss0}--\ref{eq:diss2} is not real dissipation following an entropy principle (it is not entirely irreversible), and it does not necessarily lead to heating, and in any case not instantaneously. What has been labeled $Q_+$ or $q_+$ in these equations is actually a turbulence {\myem production} rate.

Both production and dissipation occur on a finite non-zero time scale, and in any given instant, actual production may exceed dissipation (or vice-versa). In ordinary hydrodynamic turbulence the dissipative time scale $\tau$ is of order the shear time scale, but in MHD turbulence it may be much larger, as I have argued elsewhere, and as seen in, {\em e.g.}, shearing-box simulations. Plus, while the ratio of the advective rate $v_R / R$ to the shear rate $R \Omega'$ is very small in the outer thin disk, it is not necessarily so in the inner thick region considered here. Altogether then, the ratio of the advective rate $v_R / R$ to the dissipative rate $\tau^{-1}$ may equal or even greatly exceed unity in these inner regions.

Regarding dissipation and heating (or the lack thereof), MHD turbulence also differs from ordinary hydrodynamic turbulence in where its energy ultimately ends up. Much has been made of the fact that MHD turbulence possesses inverse cascades and conserved quantities that ordinary hydrodynamic turbulence does not, but there are other important differences as well. MHD turbulence in a disk creates hoop-stresses completely independently of any inverse cascades or organizing principle; a tangled field plus background shear is sufficient. As mentioned above, these hoop-stresses do real work; this represents a non-dissipative loss channel to the turbulence energy budget. In addition, turbulent energy injected into the field may be buoyantly lost vertically through the Parker instability. The field so lost to a hot, lower-beta region does not just reconnect or evaporate away, but --- depending on the radial gradient of magnetic pressure --- may snap radially towards the central axis of the system. Both of these two loss channels for turbulent energy may contribute to powering, confining and possibly collimating an outflow. 

I suggest then that protostellar jets originate in high accretion-rate systems in which the accretion flow embeds the protostar. The jets will carry away some nonzero fraction of the angular momentum of accretion but the overwhelming sink of angular momentum is the distant regions of the disk. The angular momentum flux of the jets will be far less than what standard MCA theory predicts. Direct turbulent (``viscous'') coupling turns the angular momentum problem into an energy problem. The bulk of the energy discussed above does not go into dissipation and heating, but, as an intermediate stage, into a turbulent, tangled toroidal magnetic field, capable of performing work. The jets are the exhaust for that free energy, explaining the missing boundary-layer luminosity that would otherwise be present in the case of dissipation and heating. The power for the jets comes from a combination of the accretion flow itself and the braking torque on the protostar. The braking torque in turn acts as a natural feedback mechanism, in that, being mediated by shear-driven turbulence, the braking torque increases or decreases if the spin of the protostar increases or decreases respectively.

The most promising venue for jets and related collimated outflow production by the mechanism described here is early (Class 0 or Class I), high accretion-rate protostars.
possibly including systems in FU~Ori state, although the relation between FU~Ori systems and jet production is tentative and not currently known. Nevertheless I argue that in any case it is no coincidence that, possibly outside of FU~Ori systems, it is precisely in young, high-$\dot M$ protostellar
systems that one tends to find the most powerful, well-collimated jets. Nor is it a coincidence that it is also in such systems that
the disk tends to be hot and geometrically thick, as opposed to the
cool, thin disks that were the original focus of magnetocentrifugal theory.
I therefore suggest that jets in such systems are not driven magnetocentrifugally.

\section{Accretion Preliminaries}
As pointed out above, it has actually long been recognized that an accretion disk can extract angular momentum from a protostar through the torques due to effective turbulent viscosity acting on differential rotation, but this result appears to have been neglected, or at least its importance not fully appreciated, with some noteworthy exceptions \citep[see {\it e.g.}][]{PopNar_ApJ_1991,Pop_1996}\footnote{I have also managed to find this argued pointedly in a graduate thesis somewhere, including an argument similar to what I present below, but I can no longer find this reference. I would be interested to hear from any readers who might know of it, so I might give the author proper credit.}.

The commonplace practice rather is to assume a zero-torque boundary condition (that is, zero torque due to effective viscosity; there is still an advective flux of angular momentum, {\it i.e.} an ``advective torque''). 
For example, \citet{Cha_2009} cites \citet{Ste_1998}, who cites \citet{RudPol_1991}, who cite \citet{BatPri_1981} who explicitly invoke a zero-torque inner disk boundary condition, citing \citet{Pri_1977}. But \citet{Pri_1977} only addresses an accretion disk that is thin {\myem by assumption} all the way down to the central star, and cites \citet{SS} and \citet{LBP}. \citet{SS} in turn adopt a zero-torque boundary condition because they are considering thin accretion onto a black hole, not a protostar. \citet{LBP} do indeed discuss boundary-layer emission at some length assuming a zero-torque boundary condition, but in their Appendix I, they explicitly make clear that an alternative scenario is possible, in which the angular velocity does not have a maximum, there is no boundary layer, there is a viscous torque on the star, and the torque provides up to two-thirds of the energy dissipated in the disk. It may be that thin-disk Schwarzschild or Kerr accretion should have a zero-torque inner boundary condition \citep{NovTh_1973} or not \citep{Sto_AA_1976,Sto_ApJ_1980}, and in general the issue for general black hole accretion has been controversial \citep[paragraph~2]{Pac_arxiv_2000}; for protostellar accretion however, there should be no controversy at all:  the zero-torque inner boundary condition is unjustifiable, particularly for stars rotating near breakup and accreting from a disk that is not geometrically thin.

Let $\Omega$ be the material angular velocity as a function of cylindrical coordinate $R$, and $\kappa$ be the Keplerian angular velocity as a function of $R$. Assume that the mass of the accretion flow is negligible relative to the mass of the central object, so that $\kappa \sim R^{-3/2}$ for $R > R_*$.
Let $\Omega_*$ and $\kappa_*$ be their corresponding values for the star at the stellar radius $R_*$, let $\omega$ be the normalized stellar angular rotation frequency $\omega \equiv \Omega_* / \kappa_*$, and for simplicity assume the star remains spherical with a well-defined $R_*$ regardless of the value of $\omega$. The accretion disk has a (variously defined) vertical geometrical thickness $H$ as a function of $R$. Let $b$ be, loosely speaking for now, the radial thickness of the substantially sub-Keplerian region. If $b$ is small compared to $R_*$, it is proper to call this a boundary-layer. If instead $b$ is not small compared to $R_*$, then the term ``boundary-layer'' is no longer appropriate. A boundary layer is necessarily thin by definition; see \citet{Sch_1955}. 

The standard argument for the existence of a boundary-layer with a zero-viscous-torque boundary condition is given succinctly by \citet{FKR}, but, ironically, the argument can be traced essentially unchanged back to Appendix I of \citet{LBP}. I say ironically, because again, the weaknesses of the argument are also discussed forthrightly in the very same appendix, and the authors note conditions under which it may fail.

The argument hinges critically upon the assumption of accretion from a thin disk; without that assumption, it falls apart.
Vertical thinness guarantees two things. First, it sets a practical upper limit on the magnitude of the turbulent
kinematic viscosity, which of course appears in the angular momentum equation. Second, it constrains the magnitude of the
sound speed and through it the 
radial pressure gradient that can be supplied to support the accretion flow in the sup-Keplerian innermost regions. A small effective turbulent kinematic viscosity, as defined in terms of the resultant effective Reynolds number based on the diameter of the star and a fiducial speed (say the Keplerian orbital speed), implies a relatively thin viscous boundary layer; a small sound speed, as compared to the notional speed of Keplerian rotation at a radial coordinate $R$, implies that radial pressure support will not be able to maintain a substantially sub-Keplerian disk rotation profile. The result is that, given that the star's angular velocity is not itself Keplerian, the angular velocity must have a maximum in a thin boundary layer, and using a simple effective viscous prescription for the stress, this yields a zero-torque condition.

For example, we can take a Shakura-Sunyaev type \citep{SS} local prescription for the effective kinematic viscosity due to turbulence, where $\nu_{\rm t}$ depends upon $R$. Then in the accretion disk, near the stellar surface,
\beq
\nu_{{\rm t},*} = \alpha_{{}_{\rm SS}} c_s H = \alpha_{{}_{\rm SS}} \kappa_* H^2
\eeq
and the effective Reynolds number ${\rm Re}_*$ for accretion onto the protostar, defined in terms of the stellar radius $R_*$ and the Keplerian rotation speed, may be written
\beq
{\rm Re}_* = \frac{\kappa_* R_*^2}{\nu_{\rm t}} = \alpha^{-1}_{{}_{\rm SS}} (H/R)_*^{-2}.
\label{eq:Re_SS}
\eeq
Take $\alpha_{{}_{\rm SS}} \lesssim 0.1$, and a disk that remains thin all the way down to the star, $(H/R)_* \lesssim 1/10$. Then ${\rm Re}_* \gtrsim 10^3$. This suggests a relatively thin $(b \ll R_*)$ viscous boundary layer.

Alternatively, we can take a global prescription from \citet{LBP}, in which $\nu_{\rm t}$ is fixed. Define $\alpha_{{}_{\rm LBP}}$ in terms of their\footnote{The symbol ${\cal R}_c$ is their notation for a critical Reynolds number. The notion, which is well-founded, is that turbulence tends to create an effective viscosity of sufficient magnitude such that, were it a real viscosity, the instability mechanism that led to the turbulence in the first place would be marginally quenched.}
${\cal R}_c$ as $\alpha_{{}_{\rm LBP}} = {\cal R}_c^{-1} \ll 1$. Note that \citet{LBP} adopt ${\cal R}_c = 10^3$. Write
\beq
\nu_{\rm t} = \alpha_{{}_{\rm LBP}} R_*^2 \Omega_*,
\eeq
and then the effective Reynolds number in terms of the stellar radius and the Keplerian rotation speed there, is just
\beq
{\rm Re}_* = (\alpha_{{}_{\rm LBP}} \omega)^{-1} = {\cal R}_c / \omega = 10^3 / \omega \ge 10^3.
\label{eq:Re_LBP}
\eeq
All other dynamical arguments regarding $b$ that are rooted in angular momentum conservation essentially reduce to an observation that ${\rm Re_*}$ as defined here is large.
One could argue whether it is better to use the star's actual rotation speed (as done in Lynden-Bell and Pringle) or the Keplerian speed to define an effective Reynolds number,
but for a star that is rotating at anything more than a trivial fraction of breakup, it really does not make much difference.

Turning to radial momentum balance, the dominant terms for slow infall are pressure support, with acceleration of order $c_s^2/b$, gravity
with acceleration $R\kappa^2$, and centrifugal acceleration with magnitude $R\Omega^2$. In the inner region of an even marginally substantially sub-Keplerian disk or flow this latter term diminishes in significance. Turbulent pressure and the dynamical support of large-scale coherent motions such as meridional circulations are smaller in magnitude than basic thermal pressure support, all motions being subsonic. Then, following closely the argument laid out in \citet{FKR},
\beq
\frac{c_s^2}{b} \simeq R_* \kappa^2_*
\eeq
and yet, vertical pressure balance in the disk, with a vertical effective gravity of order $H \kappa^2$, suggests that just outside the boundary layer,
\beq
c_s = H \kappa_*.
\eeq
The well-known result then is that $H$ is the geometric mean of $b$ and $R$,
\beq
b R_* \simeq H^2
\label{eq:BLthickness}
\eeq
so that again, $(H/R)_* \lesssim 1/10$ is consistent with
\beq
b \ll H \ll R_*
\label{eq:A}
\eeq
(where, say, $\ll$ here indicates at least an order of magnitude). Call eq.~(\ref{eq:A}) Assertion A.

Since realistically $\Omega_* < \kappa_*$, it is then (the argument goes) suggested that
there must be some radius $R_p > R_*$ at which the angular velocity reaches its peak value. Call this Assertion B. Using primes to denote derivatives with respect to $R$,
the boundary layer then corresponds to the region between $R_*$ and $R_p$, where
\beq
R_p = R_* + b, \ \ \ b \ll R_*, \ \ \ \Omega'(R_p) = 0.
\label{eq:BL}
\eeq
There is negligible practical difference between this definition of $b$ and alternative definitions.

Next, it is pointed out that, given
such a $R_p$, the effective {\myem viscous} flux of angular momentum (due to turbulence) across the $R=R_p$ surface must vanish. That is a simple result of the prescription tucked into eq.~(\ref{eq:diss1}) that 
\beq
W_{\langle R \phi \rangle} = \nu_t R \Omega'.
\label{eq:disk_visc_stress}
\eeq
Call this Assertion C.
The result is the zero-torque boundary condition. This fixes the integration constant $\CL = 1$ for the total radial angular momentum flux in a steady disk (see section~\ref{ss:kep}) 
and leads to familiar expressions such as
\beq
\nu_{\rm t} \Sigma = \frac{\dot M}{3 \pi} \left[ 1 - \sqrt{\frac{R_*}{R}} \right].
\eeq
Note that $\CL=1$ is equivalent to setting Novikov-Thorne ${\cal Q}=0$ at $R_*$ \citep{NovTh_1973}; this leads to divergence in $v_R$ and the vanishing of $\Sigma$ at $R_*$. Relaxation of the zero-torque boundary condition addresses these problems; again, see \citet{Sto_AA_1976,Sto_ApJ_1980}.

Applied to protostars uncritically, the entire argument may not be circular, but it certainly does have a whiff of it. 
The construction assumes a thin disk to begin with. Along with this, it also assumes that there is a clear and unambiguous distinction between the star, which is supported radially by pressure, and the disk, which is not. If we imagine a thick disk that extends all the way down to (and envelopes) the star, as in fig.~({\ref{fig:2D_recirc}}), then the situation changes, particularly when the (normalized) rotation speed of the protostar, $\omega$, instead of being vanishingly small, is substantial, say of the order of, say, $20$--$30\%$ or more as discussed  above. 
The previously convenient fiction of a clearly defined ``star'' and ``disk'' now becomes dangerously misleading.
Equation~({\ref{eq:Re_SS}}) says ${\rm Re}_* \gtrsim 10$, which is not enough to assure us a thin boundary layer. Equation~(\ref{eq:Re_LBP}) is stronger, but stands on a weaker theoretical foundation. 
And equation~({\ref{eq:BLthickness}}), given the assumptions leading to it, 
really tells us nothing at all. 

Moreover, when the disk is no longer thin, Assertion A no longer holds, and with it falls Assertion B, and possibly Assertion C as well.
In the usual argument, both B and C must hold to fix the integration constant for the angular momentum fluid equation, {\myem i.e.} the total radial angular momentum transport in the disk (advective plus viscous).
If either fails, the total flux of angular momentum in the disk is no longer thereby constrained.

Assertion C may be violated because the effective
``viscosity'' and momentum transport due to turbulence and other unresolved velocity and magnetic fields need not follow
Newton's hypothesis of linearity in the local rate-of-shear, in part because unlike real viscosity, there is
not necessarily a large separation of scales between the integral scale and the scale of structures that contribute
significantly to turbulent viscosity, particularly when the disk is no longer 
thin. The Boussinesq hypothesis\footnote{That is, turbulence transports momentum like an effective viscosity.} is, after all, only a model. 

Much more importantly though, Assertion B above is particularly
dubious because once Assertion A fails, then in a pure kinematic sense, nothing compels us to accept (\ref{eq:BL}), or even that $R_p$ even exists. In principle, the peak angular velocity could be found anywhere interior to the star, including the core if not the dead center $R=0$ outright.  
For simplicity in this paper I only consider the central star to undergo simple solid body rotation, rather than the other extreme of core-envelope (as in stellar envelope) decoupling, but even so,
the peak angular velocity may not be found interior to the disk, but in the star itself.
That is, I assume
{\myem there is no $R_p$ within the disk at which $\Omega'=0$}, and therefore relations (\ref{eq:BL}) do not hold.

\begin{figure}
 \centering
 \includegraphics[width=100mm]{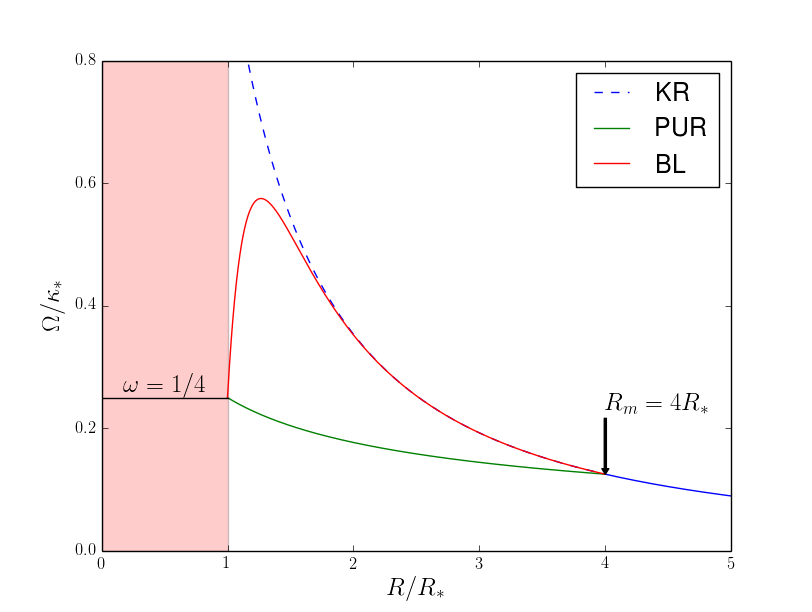}
  \caption{A central object that is rotating above or below Keplerian rotation (KR) can be embedded in a thick disk or envelope that matches to a Keplerian rotation law at some $R_m$, without possessing interior to it a peak in $\Omega$ where $\Omega'=0$ as in the standard boundary-layer form (BL). Shown is a power-law under-rotation (PUR) for $\omega = 1/4$ and $q = 1/2$, yielding $R_m = 4 R_*$.}
  \label{fig:powerlaw}
\end{figure}

For example, in \cite{Will_astroph_2001} I assumed, as I do here, that in the thick part of the flow, in the disk plane (i.e. $z=0$), the angular velocity varied as a power law with cylindrical radial distance $R$ with some exponent $-q$. This extended out to match the Keplerian thin disk at the outermost part of the thick flow, at the radius denoted $R_m$ ($m$ for ``match''), or $R_0 r_m$ in the notation of \cite{\W01}:
\beq
\Omega = 
\begin{cases} \Omega_* (R/R_*)^{-q} &\text{if $R \le R_m$,}
\\
\kappa &\text{if $R > R_m$.}
\end{cases}
\label{eq:match}
\eeq
where $\kappa = \sqrt{GM_*/R^3}$ is the usual Keplerian angular velocity. 
 See 
fig.\ {\ref{fig:powerlaw}}.
The match radius $R_m$ and exponent $q$ must satisfy
\beq
(R_m/R_*)^{q-3/2}=\omega
\eeq
A second relation would be needed to fix $q$ and $R_m$ uniquely, but that was not addressed in \cite{\W01}.

Consequences of the breakdown of relations (\ref{eq:BL}) will be explored below. The immediate implication is that the disk can couple viscously to the protostar, allowing the protostar to become a source of both mechanical energy (the energy of rotation) and angular momentum, and allowing the disk to act as a sink of that angular momentum. The amount of power available due to this coupling is, again, substantial, and possibly much more even than the notional boundary-layer power expressed in eq.~(\ref{eq:BLlum}), reducing the demands on the efficiency of the jet-production process.

Conceptually, I imagine a situation as shown, again, in fig.\ {\ref{fig:2D_recirc}}. In the inner thick region of the accretion flow, the accretion still mainly takes the form of a disk, but it is flanked above and below by confining material in recirculation zones that act as reservoirs of mass, energy, and magnetic helicity. I suspect these regions to recirculate because I expect that the magnetic stresses due to viscous coupling to the protostar will tend to generate clockwise meridional vorticity (as drawn) that baroclinicity can not counterbalance, as again discussed in \cite{\W01},
but ultimately this is just a hypothesis. In the simple treatment given here, I do not consider any coupling of conserved quantities between disk and recirculation zone. They do not exchange mass or angular momentum. The recirculation zones provide a tamper that limits radiative cooling and allows the pressure in the central plane of the disk to rise, and the turbulent viscosity to rise as well.

\section{Hydrodynamics Preliminaries}
First, I review the energetics of viscosity and viscous dissipation in an ordinary viscous 
incompressible fluid. Later, I discuss important ways in which turbulence in accretion behaves differently from a simple viscosity.

Let $\sigma$ be the viscous stress tensor, defined here as the Cauchy stress minus that part of the normal stress that is due purely to thermal pressure.\footnote{Note that this does not necessarily imply ${\rm Tr}(\sigma)=0$. For a simple incompressible Newtonian viscous fluid this will hold, but it will not hold later when these relations are carried over by analogy to the turbulent stress in accretion.} 
For a given scalar or vector $x$, let $D_t(x) = \partial_t x + {\bf v} \cdot \del x$.
The momentum equation is
\beq
\rho D_t {\bf v} = - {\bf \nabla} P + {\bf \nabla} \cdot \sigma
\label{eq:momentum}
\eeq
The stress, or equivalently the viscous momentum flux, leads to a force density ${\bf f} = \del \cdot \sigma$, as well as an energy transport due to the viscous energy flux vector $\boldsymbol{\cal f}$ where
\beq
\boldsymbol{\cal f}  = -{\bf v} \cdot \sigma.
\eeq
The divergence of (minus) this flux, $-{\bf \nabla} \cdot \boldsymbol{\cal f}$, represents the local rate ${\cal r}$ at which the energy of the fluid is increased by the action of viscous stresses, so that,
{\it e.g.} \cite{LL6} eq.~49.2,
\beq
\partial_t \left( \half \rho v^2 + \rho \sie \right) = - \del \cdot \left[ \rho {\bf v} \left(\half v^2 + w\right) - {\bf v} \cdot \sigma - \kappa_T \del T \right] 
\eeq
{\it i.e.}, in terms of the advective derivative of the total energy,
\beq
D_t\left(\frac{1}{2} v^2 + \sie \right) = -\frac{1}{\rho} \del \cdot \left( {\bf v}P - {\bf v} \cdot \sigma - \kappa_T \del T \right), 
\eeq
or in terms of the total (stagnation) enthalpy,
\beq
D_t\left(\frac{1}{2} v^2 + w \right) = \frac{1}{\rho} \partial_t P + \frac{1}{\rho} \del \cdot \left( {\bf v} \cdot \sigma + \kappa_T \del T \right),
\eeq
where $\sie$ is the specific internal energy, $w$ is the specific enthalpy, and $\kappa_T$ is the thermal conductivity.
However, there are two components to this divergence:
\beq
{\cal r} = -{\bf \nabla} \cdot \boldsymbol{\cal f} = \nabla \cdot \left( {\bf v} \cdot \sigma \right) = \underbrace{{\bf \nabla v} : \sigma }_{\text{heat}\ (+)} + \underbrace{{\bf v} \cdot ({\bf \nabla}\cdot \sigma)}_{\text{work}\ (\pm)} = {\cal q} + {\cal w}.
\label{eq:f_split}
\eeq
The first represents heating, increasing the local entropy, and in keeping with that it is generally positive-definite. (It it sometimes called the dissipation function, although that term has multiple meanings unfortunately.) The general equation of heat transfer (with specific entropy ${\cal s}$) is then
\beq
\rho T D_t {\cal s} = \del {\bf v} : \sigma +  \del \cdot \left( \kappa_T \del T \right).
\label{eq:heat_xfer}
\eeq
The second term in eqn.~(\ref{eq:f_split}) represents work in the form of bulk motion; the fluid here being incompressible, there is no $PdV$ work in any case: 
\beq
\rho D_t\left(v^2 / 2\right) = - {\bf v} \cdot \del P + {\bf v} \cdot (\del \cdot \sigma).
\eeq
This work can be either positive or negative. It is frequently convenient to re-write the kinetic energy equation as
\beq 
\rho D_t\left(v^2 / 2\right) =  -{\bf v} \cdot \del P + \del \cdot \left( {\bf v} \cdot \sigma \right) - \del {\bf v} : \sigma, 
\label{eq:ke}
\eeq 
so that the dissipation $\nabla {\bf v} : \sigma$ appears as a sink in the kinetic energy equation~(\ref{eq:ke}) and a source in the heat transfer equation~(\ref{eq:heat_xfer}).

In the case of accretion, we would not only need to consider compressibility, but more importantly, the energy equations above would need to be modified to include gravity, not to mention radiative losses and so forth. Let us ignore these for the moment however, and focus on turbulent viscosity.

Applied to a standard thin accretion disk with $v_{\langle R \rangle} \ll v_{\langle \phi \rangle}$, where now $\sigma$ is replaced by the turbulent stress tensor $W$ (Reynolds stress $R$ plus turbulent Maxwell stress ${\cal M}$) and ${\bf v}$ is the bulk (mean) flow, the two terms correspond in the following way, in the usual practice.
The dissipation function results in local heating as expressed in eqns.~(\ref{eq:diss0}--\ref{eq:diss2}), and is positive. Given eq.~(\ref{eq:disk_visc_stress}), the 
the work due to viscous stress acting on azimuthal rotation (integrated vertically) is
\beq
{\cal w}_a = v_{\langle \phi \rangle} \left( \del \cdot W \right)_{\langle \phi \rangle}  \ \ \ \ \ \ {\cal W}_a = \int_{\infty}^{\infty} {\cal w}_a\ dz = \frac{\Omega}{R} ( \nu_t \Sigma_d R^3 \Omega' )'
\label{eq:work_simple}
\eeq
and is negative, reflecting the fact that turbulence extracts mechanical energy (kinetic + potential) from the disk material.  Combined then, the disk extracts mechanical energy from accreting matter, and converts it to heat and ultimately radiation.

These two results (for heating and for work) share a key hidden assumption that may not actually hold in the inner, thick regions of accretion. Both ignore the relaxation time of the turbulence.

First, the turbulent ``dissipation function'' is not real dissipation and does not necessarily result in heating, as pointed out above. It is instead a turbulence production (or pumping) rate ${\cal p}$, and is the rate of energy going {\it into} turbulence, not the rate of energy coming {\it out} (via thermal dissipation, either by molecular viscosity or ohmic dissipation):
\beq
{\cal p} = (\del {\bf v}):W\ \ \ \ \ \ {\cal P} = \int_{-\infty}^{\infty} {\cal p}\ dz = \nu_t \Sigma R^2 (\Omega')^2.
\label{eq:heat_simple}
\eeq
We now replace eq.~(\ref{eq:f_split}) with ${\cal r} = {\cal p} + {\cal w}$, and there is not necessarily any assurance that ${\cal p} = {\cal q}$. Following a Lagrangian fluid particle, the two quantities will tend to equilibrate on the energetic relaxation time $\tau$ for the turbulence, but as the pumping ${\cal p}$ is steadily increasing as a fluid particle in-spirals, it is not guaranteed that the dissipation (or heating) ${\cal q}$ ever catches up.

Second, regarding the mechanical power, the result expressed in eq.~({\ref{eq:work_simple}) ignores the other components of the stress tensor. There are two additional contributions to the mechanical work due to turbulence, due to the normal components $W_{\langle R R \rangle}$ and $W_{\langle \phi \phi \rangle}$. The azimuthal component or hoop-stress $W_{\langle \phi \phi \rangle}$ in particular grows as the stress relaxation time $s$ grows, and both $W_{\langle R R \rangle}$ and $W_{\langle \phi \phi \rangle}$ contribute to the $R$-component of the divergence of the stress, $\left(\del W \right)_{\langle R \rangle}$. Since $v_{\langle R \rangle} \neq 0$, the mechanical work due to these stresses acting on radial motion, ${\cal w}_R$, is also not zero:
\beq
{\cal w}_R = v_{\langle R \rangle} ( \del \cdot W )_{\langle R \rangle},
\eeq
where
\beq
(\del \cdot W )_{\langle R \rangle} = \partial_R W_{\langle R R \rangle}  -\frac{W_{\langle \phi \phi \rangle} - W_{\langle R R \rangle}}{R}.
\eeq
The first of these last two terms, $\partial_R W_{\langle R R \rangle}$, is generally positive, reflecting more or less\footnote{A precise statement depends not just on an agreed definition of turbulent pressure, but also on the magnitude of the normal stress component $W_{\langle z z \rangle}$.} the radially-outward force due to turbulent pressure, including magnetic pressure. The second, $-(W_{\langle \phi \phi \rangle} - W_{\langle R R \rangle})/R$, is negative in the case of MRI-driven MHD turbulence, reflecting the inward-directed force due to the magnetic hoop stresses. The divergence is written in the form above to highlight the significance of the normal stress difference $W_{\langle \phi \phi \rangle} - W_{\langle R R \rangle}$.

This normal stress difference and the resultant inwards force, like the production term, is related to the finite non-zero relaxation time of the turbulence, but the connection is perhaps less obvious. The relevant relaxation time scale is not the energetic time scale $\tau$ but the stress relaxation (or isotropization) time scale $s$. This is explored in greater detail in the case of purely hydrodynamic turbulence in the appendix. However, for our purposes, it is sufficient to assume that $\tau$ and $s$ are one and the same, and I will just use the symbol $s$ for both going forward.

In an ordinary thin disk, ${\cal p}$ is positive-definite, and ${\cal w}$ is negative-definite. As has long been appreciated, the total rate ${\cal r}$ of energy going into the mean flow due to the turbulent stresses, including heating, can locally be either positive or negative. What has not been appreciated however is, again, that the energy that goes into turbulence might not come out in the form of heat, but through work, in the form of a positive contribution to ${\cal w}$ in the inner thick regions, approaching in magnitude the (negative) work done by reducing the angular momentum of the gas, so that the turbulence can actually do work on the gas at the same time it is robbing it of its angular momentum. To see this, we must discuss stress anisotropy, the energetics of turbulence, and their connection to relaxation in turbulence models. With sincere apologies to the reader, I will now switch to Cartesian index notation to avoid ambiguities that may unfortunately arise with the otherwise cleaner vector notation used above.


\begin{deluxetable}{llrr} 
 \tablecaption{The turbulent viscous analogy and its limits. On the turbulence side, turbulent pressure is not separated out from the total turbulent stress, as this involves some subtle issues that require further elaboration; turbulent pressure will become important however, later, particularly for its suggested role in the vertical acceleration of an outflow. Note that the total specific turbulent energy $k$ (including both kinetic and magnetic) is analogous to the internal energy $e$, but of course for a perfect fluid (on the left), $e$ is proportional to $T$. The symbol $W$ represents the turbulent stress tensor. Without compressibility, there is no distinction between the internal energy and heating equations on the left, and so the analogous equation for turbulent kinetic energy is broken out on a separate line. The ultimate line item, heating of the fluid through actual turbulent dissipation, is a sink of energy that has no analogy for a viscous fluid. Regarding the turbulent kinetic energy, I have written the loss term as ${\cal v}$. In laboratory turbulence, this is just dissipation $\rho \epsilon$, but for MHD turbulence in disks, there are other loss channels besides dissipation; see the text. I have tried to avoid subscripts $v$ for viscous and roman $\rm t$ for turbulent (italic $t$ is for time) except where necessary. In most cases it should be clear from context.}
 \tablehead{ \colhead{} & \colhead{ viscous fluid } & \colhead{ turbulent fluid } & \colhead{} }
 \startdata
conductive heating   & $ q_c = \nabla \cdot \left(\kappa_T \del T \right) $ & ${\cal d} = \del \cdot \left( \nu_{\rm t} \del {\cal k} \right) $ &turbulent self-diffusion \\
viscous energy flux  & $ \boldsymbol{\cal f} = -{\bf v} \cdot \sigma $ & $ \boldsymbol{\cal f} = - {\bar {\bf v}} \cdot W $ & turbulent energy flux \\
viscous energy deposition & $ {\cal r} = - \del \cdot \boldsymbol{\cal f} = \del \cdot \left( {\bf v} \cdot \sigma \right) = {\cal w} + {\cal q}$ & $  {\cal r} = - \del \cdot \boldsymbol{\cal f} = \del \cdot \left( {\bar {\bf v}} \cdot W \right) = {\cal w} + {\cal p}$ & turbulent energy deposition \\
viscous work    & ${\cal w} = {\bf v} \cdot \left( \del \cdot \sigma \right) $ & ${\cal w} = {\bar {\bf v}} \cdot \left( \del \cdot W \right) $ & turbulent work (on bulk flow) \\
viscous heating & ${\cal q}_v = \del {\bf v} : \sigma $ & $ {\cal p} = \del {\bar {\bf v}} : W $ & turbulent production \\
 kinetic energy & $ \rho D_t \left(v^2 / 2 \right) + {\bf v} \cdot \del P = {\cal w} = {\cal r} - {\cal q} $ & $  \rho D_t \left({\bar v}^2 / 2 \right) + {\bar {\bf v}} \cdot \del P = {\cal w} = {\cal r} - {\cal p}$ & bulk kinetic energy \\
internal energy & $ \rho D_t \sie = \rho T D_t{\cal s} - P \left( \del \cdot {\bf v} \right)$ & ---  &   \\
 heating        & $ \rho T D_t {\cal s} = q_v + q_c $ & --- & \\
                & ---                                             & $ D_t {\cal k} = {\cal p} + {\cal d} - {\cal v} $ & turbulent energy density \\
                &                               ---                                   & $ \rho T D_t {\cal s} = \rho \epsilon + \del \cdot \left( \kappa_{\rm t} \del T \right) $ & heating \\
 \enddata
\end{deluxetable}


\section{Energetics in MHD Turbulence with Zero Mean Field}

For MHD turbulence, let us perform a Reynolds decomposition, as done for purely hydrodynamic turbulence in Appendix I. To simplify matters, and in keeping with my previous work, let us assume that the mean field is {\it zero}. For the sake of avoiding factors of $4\pi$, I adopt Heaviside-Lorentz units for the magnetic field. The turbulent Faraday tension is
\beq
M_{ij} = \overline{B'_i B'_j }
\eeq
which may also be written $\overline{ \rho u'_i u'_j }$ where $u$ is the Alfv\'en velocity, highlighting the similarity to the Reynolds stress.
The turbulent Maxwell stress is
\beq
{\cal M}_{ij} = M_{ij} - \half M_{kk} \delta_{ij} = M_{ij} - P_m \delta_{ij}
\eeq
where $P_m$ is the magnetic pressure.
The full turbulent stress tensor is formed from the Reynolds stress and the turbulent Maxwell stress
\beq
W_{ij} = R_{ij} + {\cal M}_{ij}.
\eeq
and the total turbulence energy density is
\beq
{\cal k} = \rho k_R + \rho k_M
\eeq
where $\rho k_R = -R_{ii}/2$ and $\rho k_M = -{\cal M}_{ii} = M_{ii}/2 = P_m$. Here I am borrowing the standard symbol $k$ from studies of hydrodynamic turbulence wehre it is used to indicate the specific turbulent energy, such as in $k$--$\epsilon$ models, in which $\epsilon$ is the rate of specific turbulent dissipation.

Supposing that the turbulent magnetic stress is much greater than the Reynolds stress, let us now make the simplifying assumption of ignoring the latter. In this section, let us then drop the over-bar for the mean flow $\bar v_i$, since there is no longer any need to distinguish between the mean flow and the fluctuations about the mean.

The momentum equation for the mean flow is
\beq
\partial_t (\rho v_i) + \partial_j ( \rho v_i v_j) = -\partial_i P + \partial_j {\cal M}_{ij}
\eeq
which again is analogous to the momentum equation, eqn.~(\ref{eq:momentum}).

The turbulent flux of energy ${\bf f}$ is:
\beq
f_i = - v_j {\cal M}_{ij},
\eeq
and analogous relations follow for the mean kinetic energy, turbulence pumping, etc.


Let us now discuss some physical models for the turbulent Faraday (and Maxwell) stress.
First, it is convenient to adopt the notation that
\beq
{\cal D}_t M_{ij} \equiv \partial_t M_{ij} + v_k \partial_k M_{ij} - (\partial_k v_i) M_{kj} - M_{ik} (\partial_k v_j) + 2 (\partial_k v_k) M_{ij}
\eeq
which is just an expression of flux-freezing. The derivative operator ${\cal D}_t$ is a modified form of the so-called upper-convected tensor derivative. It reflects the familiar flux-freezing in ideal MHD expressed in a tensor form rather than a vector form; perfect flux-freezing is that
\beq
{\cal D}_t M_{ij} = 0.
\eeq
The simplest model is a Maxwell model for the Faraday stress. A Maxwell viscoelastic model is a model with a single relaxation time, dissipation proportional to the stress, and source proportional to an ordinary Stokes viscous term (note $\mu_{\rm t} = \rho \nu_{\rm t}$; the turbulent bulk viscosity $\zeta_{\rm t}$ is only mentioned here for formal completeness but it is not used elsewhere):
\beq
s_{{\cal q}} {\cal D}_t M_{ij} + M_{ij} = \mu_{\rm t} \left[ \partial_i v_j + \partial_j v_i - \frac{2}{3}(\partial_k v_k)\delta_{ij}\right] + \zeta_{\rm t} (\partial_k v_k) \delta_{ij}. 
\eeq
I will call this the MMF model (for Maxwell Model for Faraday stress). The subscript ${\cal q}$ for the stress relaxation time scale $s_{{\cal q}}$ stands for heating. The reason for this will become clear later.
For incompressible flows the model can be written more simply as
\beq
s_{\cal q} {\cal D}_t M_{ij} + M_{ij} = 2 \mu_{\rm t} S_{ij}
\eeq
where $S_{ij}$ is the symmetrized velocity gradient, $2S_{ij} = \partial_i v_j + \partial_j v_i$.

\citet{Ogil_MNRAS_2001} adopts a Maxwell model for the full turbulent stress $W_{ij}$, not the Faraday stress $M_{ij}$. That is a good model for his purposes, but it can lead to problems here if we assume that the turbulence is dominated by magnetic fields so that the Reynolds stress is negligible, $W_{ij} \simeq {\cal M}_{ij}$, because then the magnetic pressure has the wrong sign.

Another simple model is what I have called the $a$-$\delta$ model, which is that
\beq
s_{\cal q} {\cal D}_t M_{ij} + M_{ij} = a \delta_{ij}
\label{eq:a-delta}
\eeq
The model says this: a statistically isotropic turbulent $B$-field is created at some rate $a/s_{\cal q}$. It is then passively advected by the mean flow; this is performed by the operator ${\cal D}_t$. Finally, it dissipates at the rate $(1/s_{\cal q})$. For steady shear, setting the off-diagonal viscous stress equal to $\mu_t \gamma$, then $a = \mu_t / s_{\cal q}$. This is another simple model that is particularly appropriate if energy is being injected in part from convective turbulence in addition to shear.

The model I will adopt going forward, for concreteness, is the MMF model. Of course, each model leads to different predictions, quantitatively speaking, and the operator ${\cal D}_t$ is rather imposing in its full glory to boot, so the reader may find her or himself skeptical of the developments to ensue. However, qualitatively speaking, the difference between the three models discussed is rather minimal, and again, the operator ${\cal D}_t$ simply expresses flux-freezing, but in a tensor formalism, rather than the usual vector formalism. The predictions of all three models are dramatically different from a simple viscous model for the turbulent stress, and relative to this, the model-to-model variations are actually surprisingly minimal; furthermore, the viscoelastic models all appear to fit actual shearing-sheet simulations of the MRI far better than a simple viscous model. See for example Table~1 of \citet{Will_ASP_2003}. The improved correspondence to simulations of the MRI exists not just in the sense that the models capture the full stress tensor better (and the shear-aligned component of it in particular), but also in the fact that the models, like the simulations, show a relaxation behavior of the turbulence, in that it takes several shear time scales for the turbulence to build, and a corresponding amount of time for it to die as well, again unlike a purely viscous model.

The local loss of turbulent energy due to turbulent dissipation leading to Ohmic (Joule) heating at the magnetic Kolmogorov scale is $\rho \epsilon$. If we neglect any turbulent thermal diffusion, then this is equal to the total local rate of heating ${\cal q}_+$. (Since I am not going to discuss cooling, I will henceforth drop the subscript, and the local heating is just ${\cal q}$.) Then
\beq
{\cal q} = \frac{\cal k}{s_{\cal q}}.
\eeq

To understand energy better, let us re-write the MMF model
\beq
D_t M_{ij} = (\partial_k v_i) M_{kj} + M_{ik}(\partial_k v_j) - \frac{1}{s_{\cal q}}M_{ij} + 2\frac{\mu_{\rm t}}{s_{\cal q}} S_{ij}
\eeq
where I assume incompressibility. Taking the trace and dividing by two we have
\beq
D_t({\cal k}) = \left[ (\partial_k v_i)M_{ki} + M_{ik}(\partial_k v_i) \right]/2 - {\cal k}/{s_{\cal q}} + 0 = {\cal p} - {\cal q} + 0
\label{eq:ghfj}
\eeq

Suppose study steady uniform linear shear at shear rate $\gamma$: let $v_x = \gamma y$, and $v_y = v_z = 0$. Here, the MMF model predicts (in notation that is hopefully self-explanatory)
\beq
M_{ij} = 
\mu_{\rm t} \gamma
\begin{pmatrix}
 2 s_{\cal q} \gamma  &  1 & 0 \\
    1    &      0      & 0 \\
    0           &      0      &    0
\end{pmatrix} \ \  \ \ \ \ 
{\cal M}_{ij} = \mu_{\rm t} \gamma
\begin{pmatrix}
 s_{\cal q} \gamma & 1 & 0 \\
 1 & - s_{\cal q} \gamma & 0 \\
 0 & 0 & -s_{\cal q} \gamma
\end{pmatrix}
\eeq
The energy density is
\beq
{\cal k} = -{\cal M}_{kk} = \half M_{kk} = P_m = \mu_t \gamma (s_{\cal q} \gamma)
\eeq
and, starting from eq.~(\ref{eq:ghfj}), 
\beq
0 = D_t {\cal k} = \mu_t \gamma^2 - \frac{\cal k}{s_{\cal q}} = {\cal p} - {\cal q} 
\eeq 
Looking at the energetics of the bulk flow, what for a viscous fluid would have been the ``heating'' term is now the turbulence production term.
The frequently-encountered combination $(s_{\cal q}\gamma)$ is related (up to a factor of $2$) to the turbulent Weissenberg number, ${\rm We}$, defined as the ratio of the normal stress difference $W_{\rm xx} - W_{\rm yy}$ to the shear stress $W_{xy}$. Fits are consistent with ${\rm We} \simeq 1.5 \text{---} 8$ for MRI-driven MHD turbulence.

Production ${\cal p}$ was a source term in eq.~({\ref{eq:ghfj}}) but a loss term in the bulk flow equations, but the bulk flow equations also have ${\cal w}$ as a potential source or sink as well. Here however ${\cal w} = 0$ and ${\cal r} = {\cal p}$. Just to confirm, looking at the bulk equations we still get (remember $W = {\cal M}$)
\beq
{\cal p} = \del {\bf v} : {\cal M} = \mu_{\rm t} \gamma^2 
\eeq
We can also easily verify that
\beq
{\cal w} = {\bf v} \cdot (\del \cdot {\cal M}) = 0
\eeq
and 
\beq
{\cal r} = \del \cdot ( {\bf v} \cdot {\cal M} ) = {\cal p} + {\cal w} = \mu_{\rm t} \gamma^2.
\eeq

\section{MMF Model for MRI-Driven MHD Turbulence in Accretion}
The situation changes in accretion, in which ${\cal w}$ is no longer zero.

First however, an important and in fact absolutely key observation to the jet physics I discuss in this paper is that in the case of MHD turbulence in a disk, there is an additional loss channel besides ohmic dissipation resulting in heating, and that is the vertical loss out of the upper and lower disk surfaces. This is due to the Parker instability, {\it i.e.} buoyancy, and Alfv\'en waves, as pointed out earlier. (An analogous situation presents in pure hydrodynamic turbulence as well, in which some power may be lost vertically due to sound waves, but this is far less important effect by comparison to buoyancy and Alfv\'en waves.) This physics allows a sort of mechanical leverage, if you will, by which a sizeable fraction of the accretion energy can be given to a small fraction of the accreting mass.

Applied to a stream tube in 2D axisymmetry ({\em i.e.} a volumetric bundle of streamlines bounded above and below by an axisymmetric stream sheet), the losses (or gains) due to buoyant transport of turbulent magnetic stress into or out of the volume can be found by performing a surface integral of the turbulent energy flux vector ${\cal r}_{\rm t}$, and taking care to include compressibility effects in it via Favre averaging instead of Reynolds averaging, etc, to capture effects of rising and falling current-carrying blobs of fluid in a density- and pressure-stratified medium. This is best left to another paper. It is impractical here. Instead, perhaps motivated by a bit of intellectual laziness\footnote{I say this because it is actually an important matter to explore this flux a bit more carefully as it will have an associated torque as well (plus helicity fluxes etc), which is important to understand for overall angular momentum balance.}
as well, I adopt an {\em ansatz} that the vertical buoyant loss appears as an additional loss channel proportional to ${\cal M}_{ij}$ upon vertical integration, complementing the ohmic loss channel. Then the total loss ${\cal V}$ ({\em verlust}) per area $R\ d\phi\ dR$ of a volume element of finite vertical extent bounded below and above by lower and upper stream surfaces $\ell$ and $u$ is 
\beq
{\cal V} = D_t^{(\rm loss)} \left( \int_{z_\ell}^{z_u}\rho k_{\rm M}\ dz \right) = \int_{z_\ell}^{z_u} {\cal q}\ dz - {\cal B}_\ell + {\cal B}_u = {\cal Q} - {\cal B}_\ell + {\cal B}_u.
\eeq
This loss ${\cal V}$ is loss of turbulent energy, some of which is lost to local heating, ${\cal Q}$, and some of which truly is lost from the volume element.
One can adopt the convenient fiction of thinking of this latter as an effective local volumetric loss rate ${\cal b}$, with the caveat that with greater care it ought to be attributed to unresolved components of the (vertical) turbulent energy flux.

Assume the buoyant rise-time of magnetic flux tubes to be of order the inverse Brunt-V\"ais\"al\"a frequency\footnote{Of course it is never so simple. Drag forces on small-diameter flux tubes can substantially reduce this rate. See \citet{KhaDud_arxiv_2017}.},  which is also of order the frequency $\Omega$, and assume the loss due to Alfv\'en waves to be roughly comparable.
That is, in general on dimensional grounds one expects the time scales of both the loss $\cal b$ and ${\cal q}$ to be of order $(\text{few}) \times \Omega^{-1}$, in the context of accretion. The loss term is
\beq
\frac{1}{s}\rho k_M = {\cal v} = {\cal q} + {\cal b} = \frac{1}{s_{\cal q}} \rho k_M + \frac{1}{s_{\cal b}} \rho k_M = \frac{a_{\cal q}}{s} \rho k_M + \frac{1-a_{\cal q}}{s}\rho k_M.
\eeq
Here $a_q$ is the heating efficiency parameter. It reflects that fraction of the loss term that goes into actual heating. The remaining fraction $(1-a_q)$ of the energy is lost to the upper (or lower) boundary of the flow, and may do work on the fluid there.

Let us now adopt cylindrical coordinates $(R,\phi,z)$, and let indices run from $R$ to $z$ in that order. Primes no longer indicate fluctuations, but go back here and henceforth to indicate differentiation with respect to $R$. Here, it is sometimes useful to go back and forth between Cartesian index notation and covariant/contravariant tensor notation.
For steady azimuthal shear, the MMF model results in a Faraday stress
\beq
M_{\langle i j \rangle} = \mu_t
\begin{pmatrix}
 0 & R \Omega' & 0 \\
 R \Omega' & 2 s \left( R \Omega' \right)^2 & 0 \\
 0 & 0 & 0 \\
\end{pmatrix}
\ \ \ \ \ 
M^{\alpha \beta} = \mu_t
\begin{pmatrix}
 0 & \Omega' & 0 \\
 \Omega' & 2 s \left( \Omega' \right)^2 & 0 \\
 0 & 0 & 0 \\
\end{pmatrix}
\eeq
from which the full Maxwell stress is ${\cal M}^{\alpha \beta} = M^{\alpha \beta} - P_{\rm mag} g^{\alpha \beta}$, and $P_{\rm mag} = \mu_t s (R \Omega')^2$, so
\beq
{\cal M}_{\langle i j \rangle} = \mu_t
\begin{pmatrix}
 -s \left( R \Omega' \right)^2 & R \Omega' & 0 \\
 R \Omega' & s \left( R \Omega' \right)^2 & 0 \\
 0 & 0 & - s \left( R \Omega' \right)^2 \\
\end{pmatrix}
\ \ \ \ \ 
{\cal M}^{\alpha \beta} = \mu_t
\begin{pmatrix}
 - s \left( R \Omega' \right)^2  & \Omega' & 0 \\
\Omega' & s \left( \Omega' \right)^2 & 0 \\
 0 & 0 & - s \left( R \Omega' \right)^2 \\
\end{pmatrix}.
\eeq

The thermal pressure is $P_{\rm th} = \beta P_{\rm mag}$ for a total radial pressure force of $-(\beta + 1) \nabla P_{\rm mag}$ for fixed $\beta$.

The turbulent magnetic energy density is
\beq
\rho k_{\rm M} = \mu_t s \left( R \Omega' \right)^2
\eeq
and
\beq
0 \simeq D_t(\rho k_{\rm M}) = {\cal p} - {\cal v}
\eeq
where
\beq
{\cal p} = {\cal v} = \mu_t \left( R \Omega' \right)^2.
\label{eq:prod1}
\eeq
resulting in a heating rate $q$ and a buoyant loss rate ${\cal b}$ of 
\beq
q = a_q \mu_t \left( R \Omega' \right)^2\ \ \ \ {\cal b} = (1-a_q) \mu_t \left( R \Omega' \right)^2.
\eeq
The force density  due to the turbulence is $\nabla \cdot {\cal M} = \nabla \cdot M - \nabla P_{\rm mag}$. The turbulent radial force density ${\cal f}_{\rm t}^R$ has a pressure term, and a hoop-stress term
\beq
{\cal f}^R = \left( \nabla \cdot {\cal M} \right)^R  = {\cal M}^{\alpha R}_{\ \ \ ;\alpha} = {\cal M}^{RR}_{\ \ \ ,R} + \frac{{\cal M}^{RR} - R^2 {\cal M}^{\phi\phi}}{R}.
\label{eq:radialforce}
\eeq
There is, in addition, a torque density
\beq
R\left( \nabla \cdot {\cal M} \right)^\phi = \frac{1}{R^2} \left( R^3 {\cal M}^{R\phi} \right)'.
\label{eq:torque}
\eeq
The overall rate of turbulent energy deposition is
\beq
{\cal r} = \left( g_{\gamma \mu} {\cal M}^{\alpha\gamma} v^\mu \right)_{;\alpha} = \frac{1}{R} \left( \mu_t R^3 \Omega \Omega' \right)',
\eeq
which is the sum of turbulence production
\beq
{\cal p} = g_{\mu \nu} v^\mu_{\ ;\alpha}{\cal M}^{\alpha \gamma} = \mu_t \left(R \Omega' \right)^2
\eeq
as above in eq.~(\ref{eq:prod1}), plus the work on the bulk fluid due to turbulence,
\beq
{\cal w} = g_{\mu \gamma}v^\mu{\cal M}^{\alpha \gamma}_{\ \ \ ;\alpha} = \frac{\Omega}{R} \left( R^3 {\cal M}^{R\phi}\right)'
\eeq
related to the torque, eq.~(\ref{eq:torque}).

For the most part these are the standard results; the exceptions are: 1) The local heating is less than the nominal value by the factor $0 < a_{\cal q} < 1$, the remainder being buoyantly lost; 2) There is a radially-outward force due to the magnetic-pressure (and thermal pressure) gradient; such forces can be ignored in thin disks but not in geometrically thick accretion; 3) There is a radially-inwards hoop stress due to stress relaxation in the turbulence. 

In steady energy balance, there must be, of course, an inwards radial drift that we have neglected, so $v^R<0$. For a purely viscous model for the turbulence, the radial force (\ref{eq:radialforce}) would be zero, but this is no longer the case here. There is an additional contribution to ${\cal w}$ then, as discussed before, and the expression above for ${\cal w}$ is just the fraction ${\cal w}_{\rm a}$ of the total work.

The additional contribution ${\cal w}_r$ to the work performed on the in-spiralling gas may be approached by breaking the radial force (\ref{eq:radialforce}) into a term due to the Faraday tension (the hoop term, ${\cal w}_h$ and a term due to the magnetic pressure (the pressure term, ${\cal w}_p$:
\begin{eqnarray}
{\cal f}^R &=& \left( \nabla \cdot M \right)^R - \nabla \cdot P_m \\
           &=& - \frac{1}{R} 2 \mu_{\rm t} s \left( R \Omega' \right)^2 - \left( \mu_{\rm t} s (R \Omega')^2 \right)' \\
\end{eqnarray}
In general then, the actual total work ${\cal w}$ will be increased (that is, it will still be negative but of less magnitude) due to this additional contribution. This will be studied further below, using the formal solution in Appendix III to a presecribed velocity field in which $|v_{\langle R \rangle}| \propto R^{-1}$. 

\section{1D Steady Disks}
Let us now consider a 1D steady disk and the angular momentum flux through it. 
I construct control volumes as shown in figure~{\ref{fig:controlvolume}}; note that I assume that any jets exist {\it above and below} the star (not the disk), 
and have cylindrical radius $R \simeq R_*$. This vastly simplifies the development below. Additionally, I do not consider any coupling directly between the star and jets, or between the recirculation zones and anything else. The only couplings considered are star-disk, and jet-disk.
We could have just as well considered only star-disk and star-jet coupling instead; at this level, the difference is really a matter of accounting, not physics.

As a matter of notation, given any conserved quantity $X$, within the disk there is a radial flux $F_X$ which is a function of $R$, and the total flux into the disk is
\beq
F_X^{({d})}\biggl|_{R=R_*} = F_X^{(1)} + F_X^{(2u)} + F_X^{(2\ell)} = F_X^{(*)} + 2 F_X^{({j})} = - \dot X_* - 2 \dot X_{j} = - \dot X_{d}
\eeq
where $j$ and $d$ indicate jet and disk, and for brevity, $F_X^{(d)}$ may just be written as $F_X$. (The outer boundary of the disk is moved to infinity.)
The reason for this notation is to serve as a reminder that in general, in the disk, a flux $F_X$ may be a function of radial co\"ordinate $R$, representing the total integrated flux through a cylindrical surface, whereas $\dot X$ refers specifically to the flux evaluated at $R_*$ between star and disk or jets and disk. 


\begin{figure}
 \centering
 \includegraphics[width=100mm]{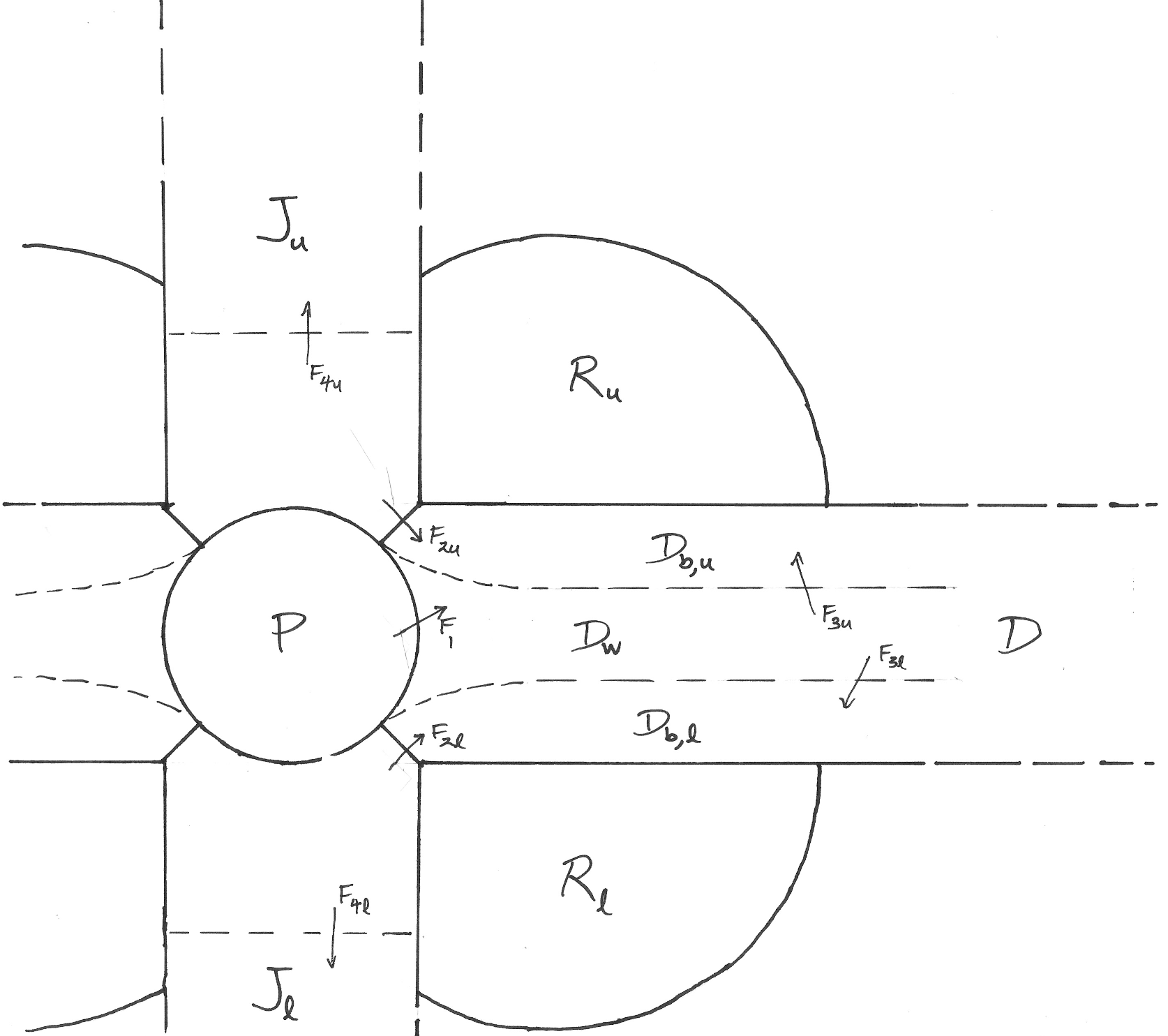}
  \caption{Conservation equations are referred to the above control volume diagram, showing a section of the 3D flow in a plane containing the axis of symmetry (vertical). Given a conserved quantity, the total flux from the disk can be broken into a flux $F_1$ from disk $D$ to protostar $P$, and a flux $2F_2$ from disk to jets $J_u$ and $J_{\ell}$, assuming $F_{2u} = F_{2\ell} = F_2$. The fluxes in the jets as determined near or past the critical surfaces 
  are $F_{3u}$ and $F_{3\ell}$. The total momentum flux (thrust) $F_j$ for a jet is necessarily determined further downstream, after the bulk of acceleration in the nozzle. There is a flow separatrix in the disk (dashed lines in $D$) separating the ``working fluid'' (material that ultimately accretes onto the protostar) stream tube $D_w$ from the upper and lower ``bypass'' fluid stream tubes $D_{b,u}$ and $D_{b,\ell}$. This is viewed as a two-stream compound-compressible channel flow. I ignore fluxes between star and jet, and between upper or lower recirculation zones $R_u$ and $R_\ell$ and anything else.}
  \label{fig:controlvolume}
\end{figure}

The fluid equations may be integrated under the simplifying assumptions already addressed above. 
Conservation of mass yields
\beq
F_M = 2 \pi R \SD \vR < 0, \ \ \ \ \Mdd = - F_M \bigg|_{R=R_*}, \ \ \ \ \frac{d}{dR} F_M = 0
\eeq
and
\beq
\dot M_* = \Mdd - 2 \Mdj = \left( 1 - \varepsilon \right) \Mdd
\label{eq:varepsilon}
\eeq
where I note again that I assume no vertical mass gain or loss, all the way down to the radius $R_*$. 
Note that the surface density $\SD$ is {\it only} that contribution to the vertically-integrated density due to the disk material, and does not include the recirculation zones. Inside the thick region, the actual surface density $\Sigma = 2 \Sigma_{\rm r} + \SD$ may be much larger than $\SD$ alone.

Empirically, it is thought a typical value of the jets' mass fraction\footnote{This is the mass accretion-ejection efficiency, also written as $\varepsilon_M$ when it is required to distinguish it from the other accretion-ejection efficiencies. See section \ref{apx:efficiencies}.}
$\varepsilon$ is in the range $2$ -- $20 \%$, leaning towards the higher number,
although the number can not be measured directly and so that inference is itself model-dependent. Arguably, $\varepsilon$ may be nearly impossible to determine in Class~0 and Class~I objects \citep{Edw_JII_2008}; veiling in particular confounds the analysis. 

Integration of the conservation of angular momentum (${L}$) equation gives:
\beq
F_{L} = -\CL \Mdd \sqrt{G M_* R_*}, \ \ \ \ \Ldd = - F_{L},
\eeq
(that is, the flux of angular momentum is independent of $R$ and normalized to a fiducial constant). For future reference, it is convenient to define a notional angular momentum flux 
\beq
\dot L_0 \equiv \Mdd \sqrt{G M_* R_*}.
\eeq
Then $F_L = - \CL \dot L_0$.

The total budget for the star is:
\beq
\dot {L}_* = \Ldd - 2 \Ldj
\label{eq:Ldot}
\eeq
where
\beq
- \Ldd = F_{L} = F_{L}^{\rm (adv)} + F_{L}^{\rm (turb)}.
\eeq
Let us assume that the specific angular momentum of the jets is that of the stellar equator, such that
\beq
2 \Ldj = 2 \Mdj \cdot R_*^2 \Omega_* = 2 \omega \Mdj R_*^2 \kappa_* =
\varepsilon \omega \Mdd R_*^2 \kappa_*.
\eeq
Regarding $F_{L}$, the advective part is
\beq
F_{L}^{\rm (adv)} = 2 \pi R^3 \vR \Omega = - R^2 \Omega \Mdd,
\label{eq:Ladv}
\eeq
whereas the turbulent part is 
\beq
F_{L}^{\rm (turb)} = - 2 \pi R^3 \Omega' \nu_t \SD.
\label{eq:Lvisc}
\eeq
Nominally, $F_{L}^{\rm (adv)} < 0$ and $F_{L}^{\rm (turb)} > 0$.


The usual practice is to set $F_{\cal L}^{\rm (turb)}$ to zero at $R_*$; that is, zero turbulent ({\em i.e.} ``viscous'') torque on the star.
For accretion to proceed without adding excess angular momentum
to the star, it has been asserted elsewhere in the literature, jets provide a sink of angular momentum, effectively allowing
$\dot {\cal L}_* \simeq 0$ in eq.~(\ref{eq:Ldot}) without any turbulent torque on the star by the disk.
This amounts to fixing an integration constant for the angular momentum equation, and in the absence of a true boundary layer, it is unwarranted. In fact, the disk is a perfectly fine sink of angular momentum as shown below.

\subsection{Keplerian Inner Envelope} \label{ss:kep}
Let us start first by examining a purely Keplerian flow before proceeding to the more realistic case of an under-rotating disk (or envelope or torus or flow) surrounding a more slowly rotating star.
It is fanciful but illustrative for this section only to imagine the central star as a solid body rotating precisely at Keplerian angular velocity at $R=R_*$, i.e. $\omega = 1$,
with a purely Keplerian disk extending all the way down to the star's surface.

\begin{figure}
\centering
\includegraphics[width=100mm]{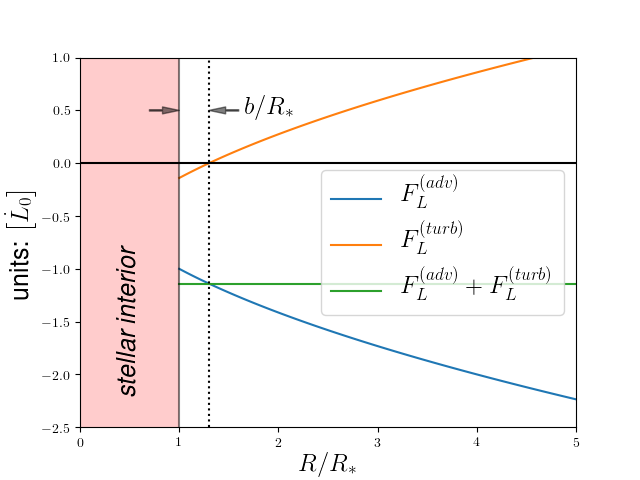}
\caption{The constant $\CL$ shifts $F_L^{\rm (turb)}$ up or down. The usual construction 
is to set $F_L^{\rm (turb)}$ to zero at $R=R_* + b$, so that $\CL = \sqrt{1+b/R_*} \simeq 1$, provided $b \ll R_*$.
Substantially decreasing $\CL$ will shift $F_L^{\rm (turb)}$ up and the total angular momentum flux up as well;
for $\CL < 0$, the total flux becomes positive, {\em i.e.} there is a net outward flux of angular momentum in the disk, removing angular momentum from the star.}
\label{fig:shifty}
\end{figure} 

In that case, eq.~({\ref{eq:Lvisc}}) reduces to
\beq
F_{L}^{\rm (turb)} = 3 \pi R^2 \kappa \nu_t \SD
\label{eq:Lvisc2}
\eeq
which may be combined with eq.~({\ref{eq:Ladv}}) to yield the familiar
\beq
\nu_t \SD = \frac{\Mdd}{3 \pi} \left[ 1 - \CL \sqrt{\frac{R_*}{R}} \right]
\label{eq:angmomK}
\eeq
and the total angular momentum of the star changes at the rate (recalling $\omega = 1$ here)
\beq
\dot {L}_* = \left( \CL - \varepsilon \right) \Mdd R^2_* \kappa_* = (\CL - \varepsilon) \dot L_0.
\eeq
Again, the total disk angular momentum flux is normalized to a fiducial value,
\beq
\CL = \frac{\dot L_{\rm (disk)}}{\dot M R_*^2 \kappa_*}.
\label{eq:CLdef}
\eeq
The conventional value is $\CL = 1$, corresponding to $F_{L}^{\rm (turb)} = 0$ at $R_p \simeq R_*$, and
$\nu_t \SD \rightarrow 0$ as $R$ approaches $R_*$.

Since in the present case however $R_p$ does not exist, we are not required to set $\CL = 1$, even for a strict viscous prescription for the turbulent stress; $\CL$ could even be negative.
All we require for $\nu_t \SD > 0$ is $\CL < 1$. For example, for $\CL < 0$ we have $F_{L}^{\rm (turb)} > -F_{L}^{\rm (adv)}$, and
the disk ``viscously'' transports away at least as much angular momentum as it advects inwards. The central star in this case gains mass but loses angular momentum, even without an outflow ($\varepsilon=0$).

Strictly speaking, we should also discuss the outer boundary condition, but the assumption here is that if there is an overall inward or outward flux of total angular momentum in the disk, then the outer disk boundary, which is otherwise effectively infinitely far away, will slowly move in or out to compensate.

For future reference, it is convenient to define a notional luminosity $\LO$ where
\beq
\LO \equiv \frac{G M_* \Mdd}{R_*} = \frac{\dot L_0^2}{R_*^2 \Mdd}
\eeq
and a notional luminosity per disk annulus $\LP$ where
\beq
\LP = \frac{G M_* \Mdd}{R^2}\ \ \ \ \ \LO = \int_{R_*}^{\infty} \LP\,dR.
\eeq



The local flux of bulk mechanical (PE + KE) energy carried by material\footnote{Radial pressure gradients are negligible here.}
in the disk is, 
with cylindrical polar coordinates $(R,\phi,z)$,
\beq
F_E^{\rm (mech)} = \frac{G M_* \dot M_d}{R} -\frac{1}{2} \dot M_d  v^2_{\langle \phi \rangle} = \frac{1}{2} \frac{G M_* \dot M_d}{R} = \frac{1}{2}R^2 \kappa^2 \dot M_d,
\eeq
giving the total flux of mechanical energy {\it to} the star, taking account of the flux to the jets, and not counting work due to turbulent ``viscous'' forces (torques), of 
\beq
\dot E_*^{\rm (mech)} = - {\cal L}_0 (1-\varepsilon)< 0
\eeq
The (negative) divergence of this radial flux,
\beq
-d_R F^{\rm(mech)} = \frac{1}{2}{\cal L}'_0 
\eeq
is equal and opposite to the work of turbulent torques (below), so that the total rate of change of mechanical energy is zero. This does not include the rate of change in specific
disk energy $-\pi G \Mdd \SD $ at a given Eulerian point $R$ due to the rate $\Mdd$ of increased mass internal to $R$, 
which is equivalent to assuming that the total mass of the disk interior to $R$ is small compared to $M_*$. This may be a bit borderline for some Class~0 sources but is probably acceptable for Class~I sources.

Turbulence energetics is a bit more complicated.

\begin{deluxetable}{lrrrr} 
 \tablecaption{Notation for key energetics quantities. Here the velocity ${\bf v}$ refers to the mean flow, so over-lines are dropped; subscripts $t$ for ``turbulent'' are also dropped. The turbulent stress tensor is $W = R + {\cal M}$, but $R=0$ by assumption. The loss of turbulence energy to buoyancy through a disk surface $\cal B$ is strictly speaking related to
 a integral of ${\bf f}\cdot dA$ over that surface, but all vertical losses through an upper or lower disk surface are lumped into $\cal B$, and ${\cal b}$ is simply the effective volumetric loss to buoyancy. The loss terms ${\cal q}$ and ${\cal b}$ are model-dependent; here, the MMF model is assumed.}
 \tablehead{ \colhead{} & \colhead{ at a point in space } & \colhead{index notation} & \colhead{ in annulus } & \colhead{ in disk } }
 \startdata
(magnetic) turbulent energy density & ${\cal e} = \rho k_{\rm M}$ & & & \\
turbulent energy flux         & $\boldsymbol{\cal f} = - {\bf v} \cdot {\cal M} $ & $- g_{\beta\gamma}v^\beta {\cal M}^{\alpha\gamma}$& $F_E^{\rm (turb)} $ & \\
turbulent energy deposition   & ${\cal r} = - \nabla \cdot {\bf f} $ & $ \left( g_{\beta\gamma}v^\beta {\cal M}^{\alpha\gamma}\right)_{;\alpha} $ & $2\pi R {\cal R}$ & $\mathbb{R}$ \\
work of turbulence on bulk flow & ${\cal w} = {\bf v}\cdot(\nabla \cdot {\cal M})$ & $g_{\mu \nu}v^\mu {\cal M}^{\alpha\gamma}_{\ \ \ ;\alpha} $ & $2\pi R {\cal W}$ & $\mathbb{W}$\\
turbulence production (``pumping'') & ${\cal p} = \nabla {\bf v} : {\cal M}$ & $g_{\mu\gamma}v^\mu_{\ ;\alpha}{\cal M}^{\alpha\gamma}$ & $2 \pi R {\cal P}$ & $\mathbb{P}$\\
total loss rate                        & ${\cal v} = {\cal q} + {\cal b}$ & & $2\pi R {\cal V}$ & $\mathbb{V}$\\
heating by turbulence (dissipation)              & ${\cal q} = a_{\rm q} \rho k_{\rm M} / s$ & &$2\pi R {\cal Q}$& $\mathbb{Q}$\\
loss to buoyancy                    & ${\cal b} = (1-a_{\rm q}) \rho k_{\rm M}/ s$ & & $2\pi R {\cal B}$& $\mathbb{B}$\\
\enddata
\end{deluxetable}

Let us begin by considering a traditional purely viscous model for the turbulent stress, and let us assume $v_{\langle R \rangle} \ll v_{\langle \phi \rangle}$ everywhere.

As before, the turbulent flux of energy can be found by examining the turbulent energy flux vector\footnote{Henceforth, except for the turbulent radial force density ${\cal f}_t^R$ as well as for the effective ``viscosity'' $\nu_t$ itself, I will drop the cumbersome subscript or superscript $t$ for turbulent, it being understood that we are no longer discussing an ideal viscous fluid.} 
${\bf f}$; in index notation, this is
$-v_\alpha \sigma^{\alpha \beta}$. In physical components in 2D ($R-\phi$) it is\footnote{This is the radial physical component $f_{\langle R \rangle}$ of ${\bf f}$, again assuming $v_{\langle R \rangle} \ll v_{\langle \phi \rangle}$; otherwise there is an additional term, $-v_{\langle R \rangle} \sigma_{\langle RR \rangle}$, where $\sigma_{\langle RR \rangle} = 2\nu_t\Sigma \partial_R v_{\langle R \rangle}$.}
$- v_{\langle \phi \rangle} \sigma_{\langle R\phi \rangle}$.
Integrated vertically and over $2\pi$ in azimuth, it is
\beq
\int f_{\langle R \rangle}\ 2\pi R\ dz = F_E^{\rm (turb)} = - 2 \pi \nu_t \SD R^3 \Omega\hspace{0.1em} \Omega' = \LO \left[ 1 - \CL \sqrt{\frac{R_*}{R}} \right]
\label{eq:EviscK}
\eeq
for a total turbulent flux of energy {\it to} the star and jet {\it from} the disk of 
\beq
\dot E_d^{\rm (turb)} = -\LO (1 - \CL) = -\Omega F_{L}^{({\rm turb})} \bigg|_{R=R_*} \le 0
\label{eq:EviscKtot}
\eeq
reflecting that for the conventional value $\CL=1$, there is no energy drawn from the star (and jet) from turbulent torques, {\it i.e.}, no
``viscous'' torque, whereas for $\CL<1$ there is a net turbulence-mediated flow of power from the star (and jet) to the disk. Note that
this is of order the energy expressed in eq.~({\ref{eq:BTorder}) as claimed.

The quantity $\dot E_d^{\rm (turb)}$ is of some importance. The notation for it follows the convention adopted above for other fluxes. However, it is also useful to look at turbulence quantities integrated over the entire disk, which I denote with chalkboard bold. The first such quantity is 
$\Rtot$, which is the total rate of turbulent energy deposition into the disk. Again, the only surfaces through which we allow a flux into or out of control volume $D$ are as shown in fig.~(\ref{fig:controlvolume}).
Then $\Rtot = - \dot E_d^{\rm (turb)}$. Furthermore, the total rate of turbulent energy deposition is equal to the sum of the total integral over the disk of turbulent production, and the total integral over the disk of the work performed by turbulence:
\beq
\Rtot = \Ptot + \Wtot.
\eeq

The negative divergence of the turbulent flux of energy is the quantity of energy $2\pi R\, {\cal R}$ being deposited (per annulus $dR$) by turbulent stresses at a radial location,
\beq
2\pi R\, {\cal R} = 2 \pi R \int {\cal r}\ dz  =  -\frac{d}{dR}  F_E^{({\rm turb})} = \LP \left[ 1 - \frac{3}{2}\CL \sqrt{\frac{R_*}{R}} \right]
\eeq
Again, in the conventional case $\CL = 1$, this quantity is negative for $R/R_* < 9/4$ and positive for $R/R_* > 9/4$, reflecting
the progressive turbulent re-distribution of the wealth of power from inner regions to outer regions of the disk, as is well-understood. Clearly, for $\CL \le 0$, it is
nowhere negative, and the disk is no longer borrowing energy from the inner regions to power the viscous dissipation in the outer regions.

The local turbulence production is (cf. eqn.~\ref{eq:heat_simple})
\beq
2 \pi R {\cal P} = 2 \pi \nu_t \Sigma R^3 \left( \Omega' \right)^2 =
\frac{3}{2}\frac{GM_*\dot M}{R^2}\left[ 1 - \CL \sqrt{\frac{R_*}{R}} \right] = \frac{3}{2} {\cal L}'_0 \left[ 1 - \CL \sqrt{\frac{R_*}{R}} \right],
\label{eq:localdissK}
\eeq
for a total power going into turbulence production in the disk of 
\beq
\Ptot = \int_{R_*}^{\infty} 2 \pi R {\cal P}\ dR = \frac{3}{2}\frac{GM_*\dot M}{R_*}\left( 1 - \frac{2}{3}\CL \right) = \frac{3}{2} {\cal L}_0 \left( 1 - \frac{2}{3}\CL \right)
\eeq
The quantity $\CL$ is not restricted by the positive-definite constraint on ${\cal P}$ (or $\Ptot$) any further than it already has been by mass conservation.

The local turbulent mechanical work due to the off-diagonal ``viscous'' stress --- that is, ignoring the normal stress --- is (cf. eqn.~\ref{eq:work_simple})
\beq
2\pi R {\cal W} = 2 \pi \Omega \left(\nu_t \Sigma R^3 \Omega' \right)' =
- \frac{1}{2}\frac{G M_* \dot M}{R^2} = -\frac{1}{2} {\cal L}'_0,
\eeq
and is independent of $\CL$; it is negative reflecting that these forces reduce, rather than increase, the energy of orbiting fluid particles. The total mechanical work of turbulence acting on differential rotation is then
\beq
\Wtot = \int_{R_*}^\infty 2 \pi R {\cal W}\ dR = - \frac{1}{2} \frac{G M_* \dot M}{R_*} = -\frac{1}{2} {\cal L}_0
\eeq
This acts to reduce the total specific mechanical energy PE + KE of fluid particles from zero (at infinity)
to $-(1/2) G M / R_*$ at the star's surface.

It can be seen that the total turbulence production in the disk is equal to minus this quantity, plus the additional turbulence-mediated flux of energy from the star, 
$\Ptot = - \Wtot + \Rtot$, this latter quantity being $0$ when $\CL=1$:
\beq
\frac{3}{2}{\cal L}_0 \left( 1 - \frac{2}{3}\CL \right) = \frac{1}{2} {\cal L}_0 +
{\cal L}_0 (1 - \CL).
\eeq

The total power going into the production of turbulent energy in the disk can be substantially larger than the nominal $(1/2) G M / R_*$. For $\CL = 0$ it is larger by a factor of 3, not coincidentally reminiscent of the famous factor of 3 discrepancy in thin-disk theory between the energy viscously dissipated in a Lagrangian annulus of the disk and the (PE + KE) energy lost by the same annulus as it moves radially inwards. In fact, almost all of this increased turbulent pumping occurs in the innermost regions of the disk so that locally, the power delivered into turbulence in these inner regions is even larger still than 3 times the
standard value, as can be seen by setting $\CL=0$ in eq.~(\ref{eq:localdissK}) and as shown in figure~(\ref{fig:kep_e_fluxes}) below.

Meanwhile the total losses $\mathbb{V}$, again as integrated over the total volume $D$ in fig.~(\ref{fig:controlvolume}), are equal to production $\mathbb{P}$ minus turbulence energy advected away into the star (and jets),
\beq
\mathbb{V} = \Ptot - \oint \rho k_{\rm M} {\bf v} \cdot dA,
\eeq
where this latter quantity is formally zero if $v_R = 0$. Also,
\beq
\mathbb{V} = a_{\cal q} \mathbb{V} + (1 - a_{\cal q}) \mathbb{V},\ \ \ a_{\cal q} \mathbb{V} = \mathbb{Q},\ \ \ (1-a_{\cal q})\mathbb{V} = \mathbb{B}.
\eeq
The total heating of the integrated volume of the disk (and thus its radiative output) is $\mathbb{Q} < \mathbb{V}$, and if we did not ignore the advection of turbulence energy into the star (and disk), we would have $\mathbb{V} < \mathbb{P}$. Of course, turbulence energy advected into the star will eventually be radiated away, but energy lost buoyantly is not (in our accounting), so in general $\mathbb{Q} \simle a_{\cal q} \mathbb{P}$. All we can say is that $a_{\cal q}$ is some fraction of unity, but in any case the total radiated power of the disk is at this point less than the conventional value, although admittedly just by the {\em ad hoc} assumption of buoyant loss.

This is the situation, in any case, regarding ${\cal R}$, ${\cal P}$ and ${\cal W}$, if we assume $v_R = 0$, but of course that is internally inconsistent, as then $\Mdd = 0$. Now if we consider small but nonzero $v_R <0$, then again as mentioned before there is an additional contribution to $\Wtot$ which includes the work of hoop stresses and radial magnetic pressure gradients.

It is convenient now, as before, to break stresses and forces down into that part due to Faraday, $F$, and that part due to the magnetic pressure, $P$. Then the turbulent radial force density ${\cal f}^R$ has two components, ${\cal f}^R = {\cal f}_F^R + {\cal f}_P^R$, where, assuming the results from the MMF model above,
\beq
{\cal f}_F^R=\nabla \cdot M  = M^{RR}_{\ \ \ ,R} + \frac{ M^{RR} - R^2 M^{\phi\phi}}{R} \simeq - R M^{\phi\phi} = - \frac{M_{\langle \phi \phi \rangle}}{R}  
\eeq
and
\beq
{\cal f}_P^R = -\nabla P_{\rm mag} = \frac{1}{2} \left( g_{\alpha\beta}M^{\alpha\beta} \right)_{,R} \simeq - \frac{1}{2}\left( R^2 M^{\phi \phi} \right)_{,R} = -\frac{1}{2} \partial_R M_{\langle \phi \phi \rangle}.
\eeq
Then, using the model results,
\beq
{\cal f}_F^R \simeq -\frac{1}{R} \left(2 \mu_t s ( R \Omega')^2 \right)
\eeq
and
\beq
{\cal f}_P^R \simeq - \left( \mu_t s (R \Omega')^2 \right)'.
\eeq
The work done by the Faraday hoop stresses in particular is then
\beq
2 \pi R {\cal W}_{RF} = \int {\cal w}_F\ dz = \frac{9}{2} \frac{s \nu_{\rm t}}{R^2} {\cal L}'_0.
\eeq
For a thick disk (and correspondingly high $\nu_{\rm t}$), and if $s\Omega$ is of order unity or larger, this is not a small quantity. Alone, it is not enough to change the sign of the Bernoulli parameter. However consider if the integral is performed over the region $D_w$. There is now a vertical flux into $D_{b,u}$ and $D_{b,\ell}$ of toroidal field, as well, with a total power of nearly $(1-a_{\cal q})\mathbb{P}$. This will also created hoop stresses, which will do comparable work on the fluid in $D_{b,u}$ and $D_{b,\ell}$, but this work is being performed on a fraction $\varepsilon$ of the total accretion mass $\dot M_d$. This is more than sufficient, it is claimed, to ultimately change the sign of the Bernoulli parameter of that gas and make it unbound. 

\begin{figure}
\centering
\includegraphics[width=100mm]{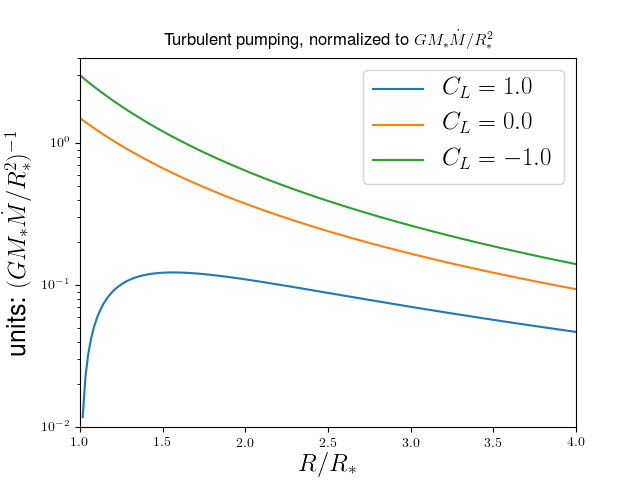}
\caption{Power extracted from star and accretion flow and injected into MHD turbulence, $2\pi R {\cal P}$,
as a function of $R$ for a steady Keplerian disk accreting onto a star notionally rotating at breakup, for different values of 
the turbulent torque on the star. Relaxation of the constraint $\CL=1$ allows the star to dump a substantially larger quantity of energy into
turbulence in the inner regions of the disk than otherwise. This energy is largely in the form of a toroidal magnetic
field, which is able to deliver energy to an outflow far more efficiently than heat energy would. Magnetic buoyancy allows much of this energy to couple to and accelerate a fraction of the mass, changing the sign of its Bernoulli parameter.}
\label{fig:kep_e_fluxes}
\end{figure}

\subsection{Non-Keplerian Inner Envelope}
There is nothing particularly special about the Keplerian disk in this regard. Of course, a sub- or super-Keplerian disk must have some kind of radial pressure support and so must not be thin, but we are not interested in solving the full problem of such disks here. Effective 1D disk equations for intermediate
``slim disks'' (Abramowicz et al) may be found by vertical integration which introduces various constants of order unity, but precise values for these constants in turn rely upon further assumptions regarding the full 2D ($R$-$z$) disk structure in the case of a truly thick disk, which can not truly be pinned down without a full $2D$ solution of course. We can still learn a fair bit by making the gross $1D$ disk approximation however. Since we are not interested in slim disks but fully thick flows, the relative uncertainty of the overall 2D structure dwarfs any precision we could bring to bear by using these constants.

Let us therefore now imagine a more realistically fast-rotating protostar, $\omega = 1/3$, with a thick inner disk or envelope\footnote{In previous work I called this inner thick accretion flow a ``disk,'' but this is a somewhat loaded word with potential connotations (esp. thinness) that I did not intend; here, I will often call this inner accretion region the ``envelope.'' That is {\em also} arguably a loaded word, but the nominal context in which one otherwise discusses an ``envelope,'' regarding protostars, is at $\gg {\rm AU}$ scales, which, it should be abundantly clear, is not what is intended here.}  with a much softer power-law dependence of $\Omega$ on $R$ than Keplerian, say $q=1/2$ or $1$, leading to $R_m = 3R_*$ or $9R_*$ respectively for the radius of this
thick inner disk, and ``sew'' this envelope onto a Keplerian thin disk for $R>R_m$. This somewhat artificial construction
will necessarily result in some complex relations and integrals
for the various quantities of interest, which should of course be taken with a grain of salt given the crudeness of the approximations used here, but counterbalancing
this is, I hope, the merit of the 
overall gross conclusions and the qualitative features of the relative magnitude of the various physical quantities.

So long as $q \neq 0$ in the envelope\footnote{Actually, even in the case of solid-body rotation $q = 0$, if not turbulent viscosity then velocity structure such as meridional circulations can lead to an effective radial transport of angular momentum so that again we do not require $\CL=1$, but in that case the transport due to such a velocity field or even turbulence can no longer be represented through a turbulent viscosity model.} we may still define $\CL$ as before, by reference to the same fiducial value for the angular momentum flux as in eq.~(\ref{eq:CLdef}), and 
\beq
\nu_t \Sigma = \frac{\dot M}{2 \pi q} \left[ 1 - \CL \frac{R_*^2 \kappa_*}{R^2 \Omega} \right]
\eeq
for this inner envelope, still keeping eq.~(\ref{eq:angmomK}) with the same $\CL$ for the outer thin disk, 
and 
eq.~({\ref{eq:Lvisc}}) still holds. Note however that $\CL$ is no longer the ratio of the total angular momentum flux to the advective flux
at the stellar surface;
that ratio is now $\omega \CL$,
and the total angular momentum of the star changes at the rate (recalling $\omega = 1$ here)
\beq
\dot {L}_* = \left( \CL - \omega \varepsilon \right) \Mdd R^2_* \kappa_* = (\CL - \omega \varepsilon) \dot L_0.
\label{eq:CLt}
\eeq

The turbulent energy flux vector becomes 
\beq
\dot E^{(visc)} = \omega^2 \frac{G M_* \dot M}{R} \left[ 1 - \frac{\CL}{\omega} \left(\frac{R_*}{R}\right)^{2-q} \right]
\label{eq:Evisc_noK}
\eeq
in the envelope but retains its value~({\ref{eq:EviscK}}) in the outer thin disk,
for a total flux of turbulent energy {\it from} the star of 
\beq
\dot E^{(visc)}_* = \frac{G M_* \dot M}{R_*} (\omega^2 - \CL \omega)
\label{eq:Evisc_noK_tot}
\eeq
which reduces to eq.~(\ref{eq:Evisc_noK_tot}) for $\omega=1$.

The local rate of turbulent energy deposition, which includes both turbulence production and work by turbulence, is
\beq
2\pi R\, {\cal R}  =  -\frac{d}{dR}  F_E^{({\rm turb})} = \dot M R \Omega_2 \left[ (2q-2) - q \frac{\CL}{\omega}\left( \frac{R_*}{R} \right)^{2-q} \right]
= \omega^2 \frac{G M_* \dot M}{R^2} \left(\frac{R}{R_*}\right)^{3-2q} \left[ (2q-2) - q \frac{\CL}{\omega}\left( \frac{R_*}{R} \right)^{2-q} \right]
\eeq

Of this, the viscous production alone is
\beq
2\pi R, {\cal P} = q \dot M R \Omega^2 \left[1 - \frac{\CL}{\omega} \left( \frac{R_*}{R} \right)^{2-q} \right] =
q \omega^2 \frac{G M_* \dot M}{R^2} \left(\frac{R}{R_*}\right)^{3-2q} \left[1 - \frac{\CL}{\omega} \left( \frac{R_*}{R} \right)^{2-q} \right] 
\eeq
for $R<R_m$; for $R\ge R_m$, eq.~(\ref{eq:localdissK}) still holds.

Meanwhile, the turbulent work in the envelope is
\beq
2 \pi R, {\cal W}_a = -(2-q) \dot M R \Omega^2
\eeq
for $R<R_m$, and $-(1/2) \dot M R \kappa^2$ in the thin disk as before.

Again, these are the results for a slow in-spiral in which $v_{\langle R \rangle} \ll R\Omega$, and for a purely viscous turbulence model (which is essentially equivalent to $s\Omega \ll 1$ in the MMF model). Figure~(\ref{fig:slow_inspiral_viscous_work}) illustrates the small correction to this due to the work of Faraday tension on the radially-inwards motion, for the case $\vartheta=0.1$ and $\hat s = s \Omega = 0.15$ from Appendix III. To satisfy overall energy conservation, ${\cal P}$ (and ${\cal R}$) also shifts by a small amount to compensate.

\begin{figure}
\centering
\includegraphics[width=100mm]{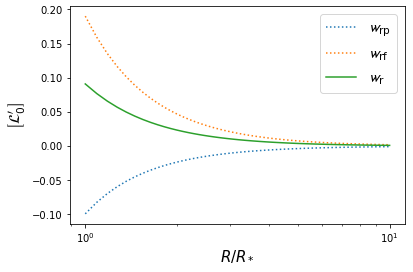}
\caption{Work of turbulence acting on radial velocity ($w_{r}$), broken down into work of Faraday stresses ($w_{rf}$ here) and work of magnetic pressure forces ($w_{rp}$) in a slow-inspiral with nearly viscous turbulence. This is for $\vartheta=0.1$ and $\hat s = 0.15$.}
\label{fig:slow_inspiral_viscous_work}
\end{figure}

The relative fraction of energy going into ${\cal W}_{RF}$ increases as $\vartheta$ and $\hat s$ increase. Consider now a much more elastic type of turbulence, let us say $s\Omega = 10.0$, ( but still with slow in-spiral, $\vartheta=0.1$); see fig.~(\ref{fig:th01_sh10}) and fig.~(\ref{fig:th01_sh10b}). Here ${\cal W}_R$ counter-balances much of the work against azimuthal motion, ${\cal W}_a$, reflecting that much of the energy robbed from the gas to allow it to accrete goes back into countering radial pressure gradients.

\begin{figure}
\centering
\includegraphics[width=100mm]{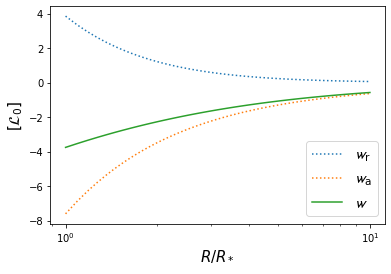}
\caption{Work of turbulence with fairly elastic turbulence (slow relaxation time compared to orbital time) showing the contributions to ${\cal W}$ due to work against azimuthal motion, ${\cal W}_a$, and work against radial motion, ${\cal W}_R$ (here, $w_r$). This is for $\vartheta=0.1$ and $\hat s = 10$.}
\label{fig:th01_sh10}
\end{figure}

\begin{figure}
\centering
\includegraphics[width=100mm]{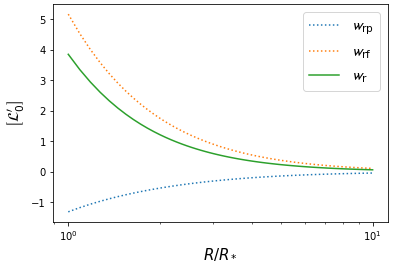}
\caption{Work of turbulence acting on radial velocity ($w_{r}$), broken down into work of Faraday stresses ($w_{rf}$ here) and work of magnetic pressure forces ($w_{rp}$) in a slow-inspiral with elastic turbulence. This is for $\vartheta=0.1$ and $\hat s = 10$.}
\label{fig:th01_sh10b}
\end{figure}

Again, there are substantial magnetic tension forces capable of performing work, and these magnetic fields can buoyantly rise from region $D_w$ to regions $D_{b,u}$ and $D_{b,\ell}$ as before, and, I argue, change the sign of the Bernoulli parameter in those streamtubes.

\section{Energy from the Star}

The energy must come from the star, which is not without its own constraints. 
The constraints on $\CL$ due to this aspect of energy conservation may be approximated as follows. Following \citet{Shu_etal_ApJ_1994} who in their section~4.1 cite \citet{James_ApJ_1964}, write the angular momentum of the star as
\beq
L_* = \omega b M_* R_*^2 \kappa_*
\eeq
where of necessity $b$ is a small fraction of unity; roughly, $b = 1/5$ decreasing to $b = 1/7$ for maximally-rotating stars. 

Stellar energy conservation depends critically upon the behavior of $\dot R_*$ with respect to $\dot M_*$.
Define the expansion factor $x_R$ and the speedup factor $x_\omega$ as
\beq
x_R \equiv \frac{ \dot R_* / \dot M_* } {R_* / M_* }, \ \ \ \ x_\omega \equiv \frac{ \dot \omega / \dot M_* }{ \omega / M_*}.
\eeq
\citet{Shu_etal_ApJ_1994} take $x_R=1$; for fixed density, for example, $x_R = 1/3$, although in principle, depending on phase of accretion, $x_R$ might even be negative. It is a purely phenomenological quantity, and is best thought of as descriptive, not prescriptive.

Then 
\beq
\frac{\dot L_*}{\dot M_*} = \left(\frac{3}{2} + \frac{1}{2} x_R + x_\omega \right) \frac{L_*}{M_*}.
\eeq
This is combined with eq.~(\ref{eq:CLt}) and eq.~(\ref{eq:varepsilon}).
For example, for a maximally-rotating ($\omega = 1$, $\dot \omega = 0$) star with $\dot R_* = 0$, then
\beq
(\CL - \varepsilon) = \frac{3}{2} b (1 - \varepsilon) 
\eeq
yielding $C_L = 1/4$ for $\varepsilon = 0.1$, and the disk carries away from the star, through turbulence, 75\% of the angular momentum that it advects to the star.
Even in this fanciful case of a star at nominal breakup and holding $R_*$ fixed,
the star does not overly ``spin up'' because turbulent torques decrease the star's angular momentum at nearly the same rate that advection increases it.

More generally,
\beq
C_L = \omega \left[ b\left(\frac{3}{2} + \frac{1}{2}x_R + x_\omega \right)\left( 1 - \varepsilon \right) + \varepsilon \right].
\eeq
Suppose $\omega = 1/3$, $b = 1/5$, $\varepsilon = 1/10$, $x_R = 1$, and $x_\omega = 0$, let us say; then $C_L = 0.153 \ll 1$. That is, about 85\% of the angular momentum advected to the protostar is carried radially back out by turbulent torques acting in the disk.

To do this, the star must yield energy to the disk, powering it (again through turbulent torques) by the quantity given in eq.~({\ref{eq:EviscKtot}}). Is this possible?

The total kinetic energy
of rotation of the star is
\beq
{\rm (K.E.)}_* = \frac{1}{2}bR_*^2\Omega_*^2 M_* = \frac{1}{2}b\, \omega^2 R_*^2\kappa_*^2 M_* = \frac{1}{2} b\, \omega^2 \frac{GM_*^2}{R_*}
\eeq
and this increases at the rate 
\beq
\frac{d}{dt} {\rm (K.E.)}_* = b\, \omega^2 (1-\varepsilon) \left( 1 - \frac{1}{2}x_R + x_\omega \right) {\cal L}_0 
\eeq
The total potential energy of the star however is
\beq
{\rm (P.E.)}_* =  - b' R_*^2\kappa_*^2 M_* = - b' R_*^2\Omega_*^2 M_* = - b' \frac{GM_*^2}{R_*}
\eeq
for some constant $b'$ and it changes by the amount
\beq
\frac{d}{dt} {\rm (P.E.)}_* = - b' (1-\varepsilon) (2 - x_R) {\cal L}_0  
\eeq
Unlike $b$ however, $b' > 1$ ({\em e.g. ca.} 1.6 for the Sun). Let us take fiducial values as before, plus $b' = 3/2$. Then
\beq
\frac{d}{dt} {\rm (P.E. + K.E.)}_* \simeq -1.34\ {\cal L}_0\ \ \ \ \ \ \ E_d^{\rm (turb)} = - 0.85\ {\cal L}_0
\eeq
Evidently, even for $x_R = 1$, there is more than enough mechanical energy for the star to provide mechanical (turbulent) energy to the disk in the quantity
given in eq.~(\ref{eq:EviscKtot}) and conserve angular momentum while doing it. This still holds all the way up to $\omega = 1$, and lower values of $x_R$ only improve the situation.
Even rotating near nominal breakup, it is only the outermost layers of the star that have a ratio of kinetic energy to potential energy approaching
a sizeable fraction of
unity in magnitude. The core dominates both the energy balance sheet and the budget, and being gravitationally bound, there is ample energy available for the star to deliver energy to the disk by turbulent stresses as given above even while keeping $\dot \omega = 0$, so long as the mass efficiency $\varepsilon$ is not too large. Thermal energy as well as energy of turbulent motions and turbulent and large-scale magnetic fields also enter into the overall accounting, and of course may also provide sinks for the gravitational binding energy released interior to the star, but these do not change the overall conclusion.

Again, as shown in fig.~(\ref{fig:kep_e_fluxes}), by reducing $\CL$ from its nominal value $\CL=1$, the energy going into either turbulence production or turbulent work in the inner regions
of the disk rises substantially.
What is the source of this energy? Ultimately the energy of course comes from gravitational binding energy, but energy like money
being fungible, 
there are many different ways to describe its flow. It should be clear in this case however that the immediate source of the energy
for this increased turbulence production and work in the inner disk is the outward flux of angular momentum from the star. However we do the accounting,
it does not make sense to say that the energy comes from the disk, if for no other reason than because the disk doesn't have enough energy to give.

\section{Schema}

\begin{figure}
\centering
\includegraphics[width=100mm]{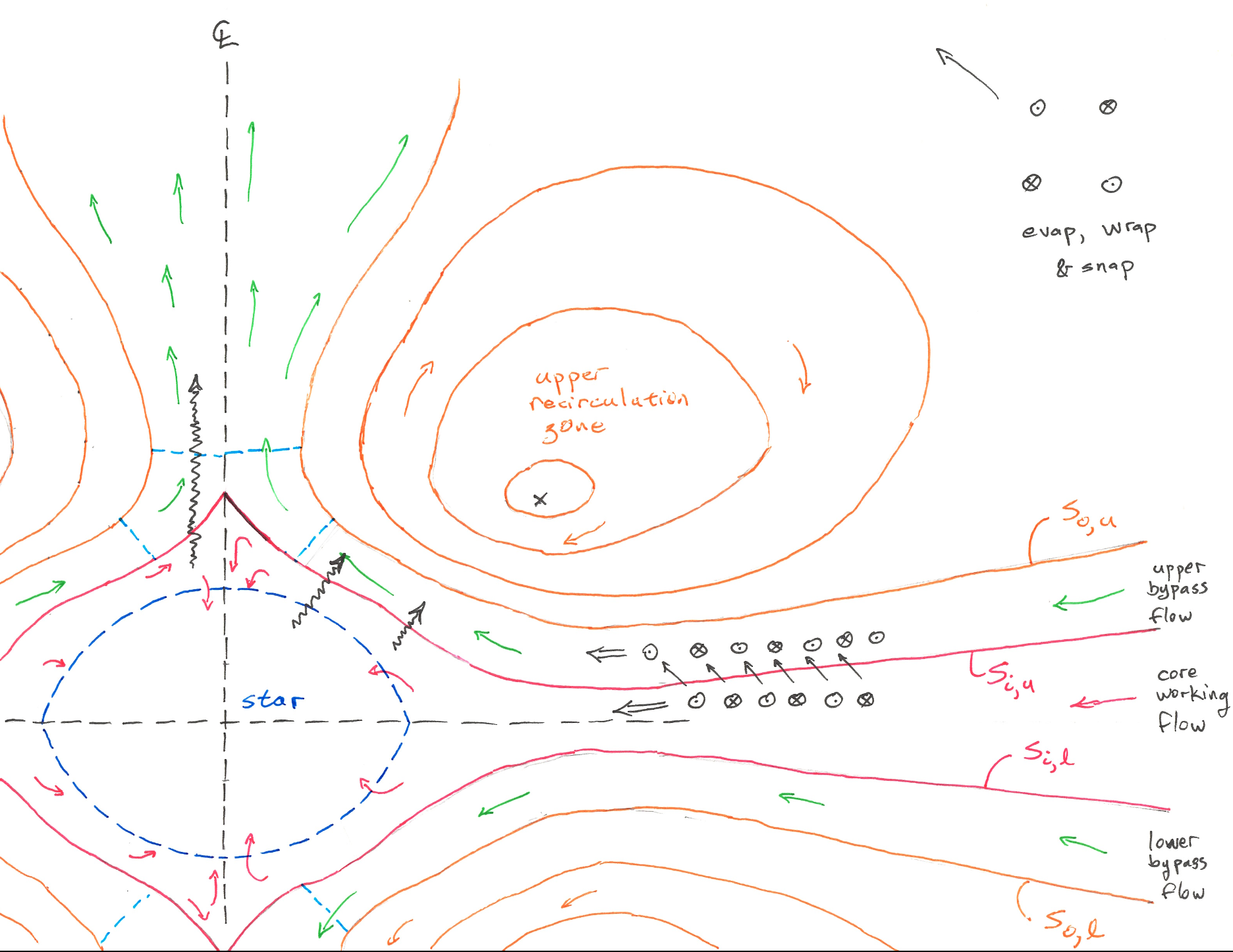}
\caption{Cartoon of schema. Turbulent hoop stresses (plus gravity) behave as a kind of compressor, powered by the ``shaft work'' of torque on the protostar conveyed by turbulent MHD stresses, compressing fluid against thermal and magnetic pressure gradients. A central ``combustion chamber'' surrounds the protostar, adding enthalpy to the bypass flows. The effective photosphere of the upper surface has the shape of an inverted Mexican hat (not shown; see fig.~\ref{fig:2D_recirc}). Critical surface is either a disk or an annulus (dashed light blue). Upper bypass fluid lies between upper inner separatrix $S_{i,u}$ and upper outer separatrix $S_{o,u}$. The stellar ``surface'' (dark blue dashed line) is somewhat arbitrary. Beyond critical surface, fluid is accelerated primarily by magnetic and thermal pressure gradients. This requires enormous enthalpy and high ($\sim 10^{6\pm0.5}\ {\rm K}$) temperatures deep at the base that would radiate copiously but for blanketing by surrounding flow. A hollow centrifugal funnel flow is avoided by an extremely high rate of turbulent angular momentum extraction from the bypass flow.}
\label{fig:schema}
\end{figure} 

The overall schema as imagined is shown in figure~(\ref{fig:schema}). The core working fluid completely envelopes the protostar and gradually settles onto it at all latitudes. Immediately above the pole is an X-point in the flow, where the inner flow separatrix meets the rotation axis (axis of symmetry). This flow separatrix separates the material that ultimately accretes onto the protostar (the core working flow) from that which does not (the bypass flow). (This separatrix was introduced in \citet{Will_arxiv_2003}; see fig.~1.1 of that paper.) There is also a separatrix dividing the bypass flow from the recirculating flow. Both separatrices have their lower counterparts.

At some range of radii between $R_*$ and $R_m$, there are recirculation zones above and below the accretion flow (i.e. the working flow and the upper and lower bypass flows). These recirculation zones serve as a tamper, providing vertical and radial pressure gradients that contribute to confining the heated accretion flow. The recirculation zones are also reservoirs of mass, energy, angular momentum, and not of least importance, magnetic helicity. Magnetic helicity is important because being conserved, it contributes to the stability and persistence of the recirculation zones. The magnetic field also provides radial (in the sense of cylindrical $R$, not spherical $r$) confinement of the outflow, as well as the recirculation zone itself. Material in these recirculation zones is near incipient $H$ ionization or recombination. The zones also shield the inner ``combustion'' region from easy direct observation. The recirculation zones are also hypothesized to be sources of meteoritic chondrules and calcium-aluminum inclusions (CAIs), but that is a subject for another paper.

Inside the working and bypass flows is a tangled, turbulent magnetic field, generated by the MRI. This is shown as alternating $\otimes$ and $\sun$. This field generates hoop-stresses, which create a radially-inwards force, indicated with double-arrows, $\Longleftarrow$. The largely toroidal field is transported vertically by buoyancy (the Parker instability) on a buoyant timescale, comparable to the advective timescale, so that the flux tubes move up at the same time they move in (slanted arrows, $\nwarrow$). One man's loss is another man's gain, and what is lost from the working fluid is gained by the bypass fluid. By this mechanism, the working flow is able to transfer mechanical energy to the bypass flow.

In addition to mechanical energy, the bypass flow is also heated. Squiggled arrows ($\leadsto$) show the thermal transport to the bypass flow both from the working flow as well as from the protostar itself, adding enthalpy to the bypass flow.

Finally, magnetic pressure from the tangled field accelerates the flow vertically through the throat. Additional enthalpy is added from what reconnection occurs in the tangled field. (In principle, $H$ recombination also adds enthalpy to the flow, but this makes little effect relatively speaking.) The flow becomes supercritical (supersonic and super-Alfv\'enic) at some height of order a few $R_*$ above the star.\footnote{Actually, from a purely kinematical perspective, there need not necessarily be a supercritical surface in terms of the fluid speed $|{\bf v}|$, rather only in terms of the poloidal speed $|{\bf v}_p|$: while it may seem physically unlikely, due especially to the  $90^\circ$ deflection of the flow in the meridional plane, in principle it can not entirely be ruled out that the flow remain supercritical (supersonic, super-Alfv\'enic) throughout.} Expansion and acceleration continue in the supercritical region beyond. Radial (in the sense of cylindrical $R$) expansion is hindered both by external hoop-stresses acting to some extent as the diverging part of a $\text{de}\,\text{Laval}$ nozzle, but the flow is also radially (again in the sense of $R$) confined and collimated by the toroidal component of its own internal tangled field.\footnote{Caution is required here (R. Blandford, personal communication, 2002). For an isolated current line, the radial inwards force due to the hoop-stress due to the toroidal field is exactly countered by the outwards force due to the magnetic pressure gradient. XXX...} 
It is speculated that this field (internal and external) mitigates the shear-induced Kelvin-Helmholtz instability that would otherwise destroy this radial confinement. 

Additional collimation is provided by tangled field that has been lost from the disk by buoyancy (``evaporation''), shown in the upper-right corner. In reality, it is highly unlikely that a flux tube would rise uniformly at all angles in azimuth around the disk at once. Rather, the flux tube can be thought of like a solar prominence, having two footprints in the disk. These footprints will generally not be at precisely the same distance $R$ from the protostar. As a result, the flux tube may experience differential stretch (if the footprint at smaller $R$ is also ahead of the other, in the sense of its rotation around the protostar), or ``wrap.'' Ultimately, the flux tube will ``snap'' inwards radially (in $R$). This motion is against the main flow which as drawn is radially outwards. Either there is a flow reversal (not shown) or there is a relative motion between the flux tubes and the background flow. This ``evap, wrap and snap'' action has not been addressed in this paper as the focus has been on the region interior to the thick recirculation zones. But, it should be expected to occur at some level, adding both to the collimation and the angular momentum flux of the outflow. Similar notions have been put forth by other researchers. Primarily, this physics is notable in the present context because such additional field wrapping may add a non-negligible quantity of angular momentum to the outflow, in which case that needs to be taken into account, both in the total angular momentum flux of jets and in the angular momentum balance in the disk outer to $R_m$.



Assume the jet is in a low-beta state, say plasma $\beta$ of order $\beta = 0.10$, so $v_A \simeq 3 c_s$. The relevant wave speed is the bulk elastic wave speed, but this is within a factor of order one of the Alfv\'en speed. Suppose the gas is fully ionized and at a temperature of, let us say, $10^5\ {\rm K}$, with a sound speed of $c_s \simeq 40\ {\rm km\ s^{-1}}$. Then the
jet speed at the critical point is
\beq
v_0 = 130\ {\rm km\ s^{-1}}\ \left( \frac{\beta}{1/10}\right)^{-1/2} \left( \frac{T_0}{10^5\ {\rm K}} \right)^{-1/2}
\eeq
and the density in the jet at the throat is of order
\beq
\rho_0 = 10^{-13}\ {\rm g\ cm^{-3}}\ \left( \frac{\varepsilon_M}{0.2} \right) \left( \frac{\beta}{1/10}\right)^{-1/2} \left(\frac{\Mdd}{10^{-8}\ M_\sun\ yr^{-1}}\right)   \left( \frac{R_*}{3\ R_\sun} \right)^{-2} \left( \frac{T_0}{10^5\ {\rm K}} \right)^{-1/2}
\eeq
or an electron number density, for the same fiducial parameters, of about $n_e \simeq 10^{11}\ {\rm cm^{-3}}$. The jet will continue to accelerate past the critical point, and of course cool adiabatically as well so that $c_s$ drops and the jet becomes highly supersonic within an order of magnitude $R_*$ or so vertical distance from the protostar, with some additional acceleration provided by the substantial enthalpy of $H$ recombination.

Given these numbers, the one-sided Bremsstrahlung luminosity might be expected to be of order
\beq
{\cal L}_X = 2 \times 10^{32}\ {\rm erg\ s^{-1}}.
\eeq
This is larger than the nominal ${\cal L}_x \sim 10^{29.5\pm0.5}\ {\rm erg \ s^{-1}}$ \citep{Gud_etal_AA_2007}, but the explanation for this is strong absorption by the partially-ionized recirculation zones. For example, the assuming a line-of-sight column density through the recirculation zone comparable to $R_* n_e $, the optical depth just from Thomson scattering alone is roughly $1.0e{-2}$. In practice, X-ray optical depths from bound-free transitions in passing through a recirculation zone will be substantially higher. Being relatively compact, the column density passing through the recirculation zone can be relatively large without requiring an inordinate mass.

Regarding missing boundary-layer luminosity, the answer I suggest is twofold. First, there is less energy deposited into heating than might be expected. Second, the volume over which this energy is deposited is much smaller than a thin boundary layer on the surface of the protostar.






\section{Discussion}
Thermodynamics provides both the simplest and most stringent constraints on any jet theory. Energetically, it would appear that if we only channel the energy of accretion back into the accreting material itself, and we assume that all inflowing mass becomes outflowing so that $\varepsilon = 1$, then the best we can do is an outflow that fizzles, meaning it fails to become energetically unbound. Clearly the mass accretion-ejection efficiency $\varepsilon$ is not unity. 
Even so, the amount of mechanical energy available in the turbulence in accretion onto a rapidly-spinning protostar --- again, mechanical, not thermal --- is actually much more than the notional accretion energy $\LO/2$. In a more realistic scenario in which $\varepsilon$ is small ($\varepsilon = 0.2$, say), first, the required overall thermodynamic efficiency of the jet-launching engine is actually less than is traditionally claimed, and second, the energy being mechanical rather than thermal in nature, the required thermodynamic efficiency is actually achievable as well.

Regarding the MRI in particular for its role in generating the MHD turbulence that is an intermediary in providing energy to the jet, to quote from \cite{Will_ASP_2003}, ``That a tangled field can have such important dynamical consequences as angular momentum transport in accretion disks leads us to ask what other consequences such a field might have. We propose that one consequence of this tangled field may be the driving of an axial outflow, {\em i.e.} a jet.'' As there appears to have been some confusion on the part of the referee for \citet{Will_astroph_2001} in this regard,  I attempted to clarify in \citet{Will_MNRAS_2005}, ``To be quite explicit, as a hypothesis, I reject the notion that \dots magnetocentrifugal mechanisms are responsible for tightly collimated jets.'' The confusion likely arose because conventionally, in MCA theory, a far-field toroidal field is invoked to provide collimation, whereas what I have discussed everywhere, consistently, is a near-field toroidal field, up to and including the very mid-plane itself. There is an issue of semantics, as a rocket engine both requires collimation in the supercritical region (the nozzle), and at or near the critical point itself, as well as in the subcritical region (the reservoir and the inlet to the nozzle). Perhaps this latter is best called ``confinement,'' in which case the title to my paper \citet{Will_MNRAS_2005} should be changed, from ``\dots jet collimation \dots'' to ``\dots jet material confinement and collimation \dots''; even so, in the case of collimation of supercritical ($v > v_A, c_s$) flow, collimation is only needed before the jet becomes highly supersonic (and super-Alfv\'enic). Beyond this point the Mach angle is small, and lateral spreading is minimal. From this perspective, in the schema presented here, collimation is mainly needed from $R_*$ to $0.1\ {\rm AU}$ distance in vertical separation from the star (for a fiducial $3\ R_\sun$ and $0.8\ M_\sun$ protostar).

Since the schema I have discussed here and from the very beginning in \citet{Will_astroph_2001} is a manifestly non-magnetocentrifugal picture of how jets might be created, it is worthwhile to discuss the grounding of the MCA theory. To date, observations of jets appear consistent with the MCA hypothesis. But consistency is not in itself enough.

At the grossest level, the appeal of MCA theory is that it solves the angular momentum problem, allowing protostars to condense and accrete by dumping excess angular momentum into a jet. This had a definite and considerable appeal before the discovery of the role of the MRI in accretion \citep{BalHaw_ApJ_1992}, but arguably less so now that the MRI has been (re-)discovered and studied intensely. The MRI does appear to have ``dead zones,'' locations where it is not active, in protostellar accretion \citep{Gam_ApJ_1996}. But these are farther out in radius ($R>0.1\ {\rm AU}$), likely intermittent, and in any case the surface layers still remain active due to cosmic-ray ionization. Some flavor of MCA may indeed ``help'' accretion proceed by allowing angular momentum to escape to infinity, but it is not at all clear --- to me, at least --- that this is required. The MRI is quite effective in doing this already, by transporting angular momentum radially outwards, not vertically, and it is fully capable of doing this all the way right down to the protostar itself. This arguably has a much simpler physical appeal, for the same reason that one opens a door at the handle, not at the hinges. Jets do not allow accretion to proceed by providing a place to dump excess angular momentum. Rather, they are a way for the system to discard excess {\em free energy}, not just from the disk, but from the star itself as it condenses.

One prediction of the MCA hypothesis, both in the standard self-similar disk-wind theory and in the X-wind theory in which the locus of initial magnetocentrifugal acceleration is near the protostar, is that the far-field material velocities are roughly of the order of the escape velocity near the protostar. That is, the asymptotic specific energy of the material is of the same magnitude but opposite sign, roughly, as it had near the star. On the one hand, this is an important constraint, and tends to argue against jets being created by the acceleration of material coming from more remote ($R \gg R_*$) regions of the accretion flow. On the other hand, it is not a terribly restrictive  constraint, as any mechanism that operates in the immediate neighborhood of the protostellar surface might be expected to eject material at similar speeds. However difficult it may be to construct a scenario that will take material near a stellar surface and make it energetically unbound, once this has been accomplished, it would require a delicate balancing act to do so without giving such material much more than a trivial fraction, at infinity, of the (negative) energy it formerly had (but with opposite sign). Besides, as discussed below, energy is not the problem. As emphasized by X-wind theory adherents as well, there is abundant mechanical energy not just from the accretion flow but from the star in the form of rotation.

The problem, again as emphasized in the introduction, rather, is one of efficiency, first and foremost thermodynamic efficiency, and secondly, mass efficiency.\footnote{That is, once you have figured out a way to give energy to an outflow, how to you find a way to give that energy only to part of $\Mdd$ and not all of it?}
Indeed, MCA theory is appealing here because it provides a theoretically thermodynamically efficient method to accelerate material. But as I have discussed above, neither is this as restrictive a constraint as might be thought. The notion that ``viscous dissipation'' in the disk can not be thermodynamically efficient at driving jets rests on a mistake: ``viscous dissipation'' as treated in standard disk theory is neither viscous nor dissipation at all, but turbulent production (pumping), leading to mechanical energy -- notably tangled, turbulent, buoyant, encircling and constricting magnetic fields -- not heat, at least not directly, and not necessarily on a time scale short compared with the advective time scale.

Nor is MCA theory without its problems. Among these are of course the problem of the origin of the large-scale poloidal field (is it created by flux conservation as part of the process of condensing of the initial gravitational collapse of a potential star-forming cloud, advected from afar, or created on-site by the disk, or by the star, or some combination?), and more importantly, the problem of on the one hand threading disk material onto the magnetic field while on the other hand ensuring that the field is strong enough to centrifugally accelerate the same material once it is so ensnared.

Another interesting problem that arises in standard MCA theories of jets and indeed with any MHD theory of jets that relies upon large-scale magnetic fields is the persistence of the conserved quantity magnetic flux. As mentioned {\em e.g.} by \citet{Hart_2008}, to quote, ``for a magnetized disk wind the field will be mostly toroidal at large distances and will scale as $r^{-1}$, while the density drops as $r^{-2}$ for a collimated jet.'' The implication is that the Alfv\'en speed remains constant, and the jet, if initially strongly magnetized at its base, should remain so at large distances. This is at odds with measurements \citep{Mors_etal_1992} of the magnetic field in the working surface of jets (as inferred from shock compression ratios as determined by spectroscopic measurements of the free electron density), which indicate that fields at large distances (e.g. $\simge 10^{4.5}\ {\rm AU}$ for HH\ 34) are not dynamically significant. This is also consistent with the observation of shocks in jets, which would otherwise not form in the presence of a dynamically-significant field. One solution as mentioned by \citet{Hart_2008}, see \citet{Hart_etal_2007}, is that the jet launching process itself is strongly time-dependent.

Acceleration by a turbulent, tangled field, as I have discussed here and elsewhere beginning with \cite{Will_astroph_2001}, offers another solution. A tangled field is not encumbered by a magnetic flux constraint. The total magnetic flux through any surface perpendicular to a stream tube in the outflow can be identically {\em zero}, at the same time that the Alfv\'en speed is large (e.g. of the same order of magnitude as the ejecta speed, larger or smaller depending on if we have yet crossed the critical surface), only it is a turbulent Alfv\'en speed corresponding to effective elastic modes \citep{GruDia_PP_1996, Sche_etal_NJP_2002, Will_NA_2004} due to the turbulent elasticity of the bulk fluid that is the relevant quantity, not the Alfv\'en speed corresponding to the mean field. In this sense we can have our cake and eat it too, by having both a large magnetic field (in magnitude) that can accelerate material though magnetic tension forces and magnetic pressure forces, and at the same time a negligible magnetic field (in terms of total flux), so that the magnetic field can reconnect and dissipate at large distances.

For example, as cited by \cite{Cabrit_2007}, \citet{Ferr_JI_2007} suggests a poloidal enclosed magnetic flux at $1\ {\rm AU}$ of $\Phi_B \simeq 10^{26}\ {\rm G\ cm^2}$, whereas we have $B_{\rm coll} \neq 0$ while $\left( \Phi_B \right)_{\rm coll} = 0$.

Another way of saying this is that in the schema I have presented, one does not care at all about closing the global circuit. No Biermann battery or the like or indeed any global current at all is required to maintain the large-scale field, as there is no such field to begin with. (Of course in reality there will be {\em some} nonzero mean field, but this does not provide a meaningful constraint to the gross picture at this point, in any case.) 

Yet another appeal of MCA is its analytical tractability. This is rooted in self-similarity (in the original theory). Symmetry is a phenomenal tool when you can use it, such as in the Sedov-Taylor blast wave theory. Nature is not necessarily so kind, however. Already it is clear that scaling invariance is broken first by the presence of the star itself (assuming, again, that jets are launched near the star), and secondly by microphysics, such as incipient ${\rm H}$ ionization and its effect upon opacities, such as in the recirculation zones I have posited. It should not be surprising at all if, as claimed here, the jet launching process is not at all respective of self-similarity.











The most stringent and direct support for the MCA hypothesis to date, rather, rests in observations of jet rotation, and indeed observations of such rotation have been widely seen as supportive of the MCA hypothesis (Cabrit, Bacciotti). In contrast, the mechanism described here tends to create jets with a much smaller specific angular momentum flux than should be expected from MCA-type models.
Concerning this, let us look at some fiducial numbers from possible observations of jet rotation. From Bacciotti et al 2007, observing DG~Tau and several other T~Tauri stars, let us take azimuthal velocities in the range $6$--$25\ {\rm km\ s^{-1}}$, at radial separations of $20$--$30\ {\rm AU}$. Taking the geometric mean of both quantities yields an observed specific angular momentum $\ell_{\rm obs}$ of
\beq
\ell_{\rm obs} = 300\ {\rm AU\ km/s}.
\eeq
In contrast, the specific angular momentum of the jet-launching process described here is much less, of order
\beq
\ell = 1.05\ {\rm AU km/s}\ \left(\frac{\omega}{0.33}\right)^{} \left( \frac{ M_*}{0.8\ M_\sun} \right)^{1/2} \left( \frac{R_*}{3\ R_\sun} \right)^{1/2}.
\eeq
However, the specific angular momentum just due to Keplerian rotation at $20$--$30\ {\rm AU}$ is of order $100$--$150\ {\rm AU\ km\ s^{-1}}.$ If such observations of rotation hold up, there remains the question of whether the material observed is the true jet core itself, or is rather surrounding entrained or evaporated material. It is not just the angular momentum of material in or around the jet that is of interest, but the flux of angular momentum in particular, and the ratio of this to the turbulent flux in the disk itself. It is claimed here that that ratio is low, whereas MCA says otherwise. Outside of observations at the sub-0.1~AU scale, this ratio is the critical discriminant and test of MCA theory.







\section{Appendix I: Accretion Ejection Efficiencies} \label{apx:efficiencies}
The discussion here is based heavily upon \citet{Cabrit_2007} with some additional comments. There are three traditional measures of the efficiency of jets, related to the three conserved quantities of mass, momentum, and energy, and to which we might add a fourth, related to angular momentum.

The first is the mass efficiency, defined in terms of that fraction of accreting matter that ends up in a jet (or outflow). It should be noted from the outset that it is not at all clear that such a quantity even exists however. If material recirculates, then there may be no clear way of saying what matter is actually accreting; furthermore, outflows may happen at various locations (e.g. radii $R$), as may inflows; should these all be counted?

Caveats aside, the definition (assuming two opposing jets that are in all other senses equal) is
\beq
\varepsilon_M \equiv \frac{2 \Mdj}{\Mdd} \simeq 0.2.
\eeq
(this and other typical values are taken from \citet{Cabrit_2007}). This quantity being of prime importance, in the main body of the text, I have simply denoted this $\varepsilon$, dropping the subscript.
Actually, it may be worth noting that $\varepsilon_M$ is usually defined in terms of the accretion rate $\dot M_{\rm acc}$, so one could say that I am assuming here that $\dot M_{\rm acc} = \Mdd > \dot M_*$; the difference between $\Mdd$ and $\dot M_*$, given other uncertainties, though, is almost certainly negligible in most if not all cases. As an amusing if somewhat strained analogy, thinking of astrophysical jets as turbofan engines, $\varepsilon_M$ is related to a quantity in turbofans known as the bypass ratio, $\varepsilon_M / (1- \varepsilon_M) $. In this sense, protostellar jets are like low-bypass turbofan engines.

Momentum (or thrust) efficiency relates the thrust to the accretion luminosity $\Lacc$. The one-sided jet thrust is $F_j = \Mdj v_j$; it is assumed that $F_j$ is taken here at a location far enough from the star that it has approached its asymptotic value (that is, gravity, thermal pressure, and magnetic pressure are no longer decelerating or accelerating the outflow; for an observer, this is a non-issue). This is then compared to the notional momentum flux $\Lacc / c$, which is the momentum that would be obtained if all the photons from $\Lacc$ were pointed in one direction like a flashlight. Then
\beq
\varepsilon_F \equiv \frac{2 \Mdj v_j}{\Lacc / c} \simeq 300.
\eeq
The fact that $\varepsilon_F$ is so large argues strongly against scattering of light from $\Lacc$ being a substantial source of the momentum of jets. This is a very large number for an ``efficiency'' however; an alternative sort of efficiency might normalize to the thermal velocity of gas at the stellar surface, or to the escape velocity. The escape velocity from the stellar surface is
\beq
v_{\rm esc}^{(*)} = 320\ {\rm km\ s}^{-1}\ \left(\frac{M_*}{0.8 M_\sun}\right)^{0.5} \left( \frac{R_*}{3R_\sun} \right)^{-0.5}.
\eeq
Defined in terms of this, an alternative thrust efficiency measure is
\beq
\varepsilon'_F \equiv \frac{ 2 \Mdj v_j}{\Lacc / v_{\rm esc}^{(*)}} = 0.32.
\eeq
This is a quantity of more theoretical interest than observational, clearly.

The energy efficiency is defined in terms of the mechanical luminosity of the jets; the one-sided mechanical jet luminosity is ${\cal L}_j = \Mdj v^2_j / 2$. For this quantity, it is important that one
determines ${\cal L}_j$ close enough to the source that there has been little entrainment of surrounding material that would otherwise reduce the jet speed (and increase its mass loading). The momentum flux of jets is generally unchanged by entrainment but the energy flux is, as it must be; entrainment requires dissipative structures in order to satisfy conservation.
Then
\beq
\varepsilon_E \equiv \frac{2 {\cal L}_j}{\Lacc} \simeq 0.15.
\eeq
In principle $\varepsilon_F$ and $\varepsilon_E$ are determined independently, but in practice this may be difficult.

The fourth and final efficiency is the angular momentum efficiency. To define this, we need a notional angular momentum flux. One choice is
\beq
\dot L_0 \equiv \Mdd \sqrt{ G M_* R_* }.
\eeq

Then
\beq
\varepsilon_{L} \equiv \frac{2 \dot {L}_j}{\dot {L}_0}.
\eeq

Under this definition, $\varepsilon_{L}$ for the scenario I have described, ignoring entrainment, is $\varepsilon_{L} = \varepsilon_M \simeq 0.2 < 1$. This ignores the ``evaporate, wrap and snap'' action of toroidal field components from $R \gg R_*$; these will carry angular momentum as well, increasing the total angular momentum flux in the outflow, and possibly in the jet itself. It does not take much entrainment to vastly inflate $\varepsilon_L$. Still, offhand one expects the efficiency $\varepsilon_L$ to be much less than it would be for MCA theories, in which the entire point is that the jet enables accretion by carrying away a substantial angular momentum flux.

\section{Appendix II: Energy and Stress Relaxation in Hydrodynamic Turbulence}

Although our interest is MHD turbulence, this section discusses relaxation time and anisotropy in the context of purely hydrodynamic turbulence. This is done to compare and contrast with MHD turbulence, and in order to provide a bit more intellectual continuity with studies of purely hydrodynamic turbulence, as it is more mature and well-developed topic than the theory of MHD turbulence. As the modeling minutiae may distract from the overall thrust of this paper however, readers should feel free to skip over some of the more weedy details presented here.

Like all crude models, closer inspection reveals finer distinctions.
There is the one hand the time scale $s$ on which the anisotropy of the turbulence stress relaxes, and
there is on the other hand the time scale $\tau$ on which the kinetic (or magnetic) energy in turbulence is ultimately dissipated by molecular viscosity or resistivity into actual heat. In previous work I have ignored the distinction, as the two time scales were essentially synonymous for the simple models I considered, and I largely focused on the stress relaxation in any case. I claimed that given a shear rate $\gamma$, the product $s\gamma$ is of order unity for hydrodynamic turbulence, but potentially much larger for MHD turbulence. (As always, I am assuming a zero mean-field in the MHD case.) In hydrodynamics, anisotropy is carried on the largest scales; small eddies contribute only an isotropic component to the stress. A key implication is that the normalized first normal stress difference for hydrodynamic turbulence is small, of order unity, whereas it is potentially much larger for MHD turbulence, as appears confirmed by shearing-sheet simulations of MRI-driven turbulence in particular.

For a simple shear in Cartesian geometry in which\footnote{In the literature on continuum mechanics and non-Newtonian fluids, the shear rate is often denoted $\dot \gamma$, but here and elsewhere I simply use $\gamma$.}
$v_x = \gamma y$ and $v_y = v_z =0$, the normalized first normal stress difference is $(\sigma_{xx} - \sigma_{yy})/\sigma_{xy} $.\footnote{It might seem preferable to characterize the stress anisotropy given by the normal stress difference in terms of the principle stresses instead. There is no added benefit of this here however, and the first normal stress difference for a steady shear is a quantity of historical and practical significance. Note that in this section I have dropped the angle brackets in tensor components as the analysis here is entirely in Cartesian coordinates and so no distinction is required between covariant, contravariant, or physical components.}
For a Newtonian viscous fluid --- and for standard theory of accretion disk turbulence and indeed any model of turbulence that adopts the Boussinesq hypothesis --- this quantity is zero. Effectively, and loosely, this quantity is related to the relaxation time of the anisotropy of the stress. For example, for purely hydrodynamic turbulence, the notion that eddies on the largest scale have a coherency time of order of the eddy-turnover time implies that the normalized first normal stress difference is of order or unity. On the other hand, it takes very many large scale eddy-turnover times (or shear times) for energy to cascade all the way down to the Kolmogorov scale and thermally dissipate. In general, it seems reasonable to expect that $\tau$ may be much greater than $s$, but certainly not much less. This is discussed further below.

Consider purely hydrodynamic turbulence in an incompressible fluid of fixed density $\rho$. In this section, I adopt Cartesian index notation, with $\delta_{ij}$ as the Kronecker delta. For this and the next section only, primes indicate fluctuations, not derivatives with respect to radial coordinate $R$.
These fluctuations correspond to a Reynolds decomposition with standard Reynolds rules of averaging. Over-lines indicate ensemble mean quantities, so that a given field variable $x$ may be written
\beq
x = \bar x + x'.
\eeq
The intermediate results presented below are well-known and the various symbols adopted, while not universal, are fairly standard. See \citet{TenLum_1972}, \cite{Pope_2000}, \cite{David_2004}, or especially \citet{Wilcox_2006}; for an excellent introduction, see ch.~3 of \citet{VerMal_1996}.

The nomenclature chosen here is that the Reynolds stress is $R_{ij} = -\rho \overline{v'_i v'_j}$, and the specific Reynolds stress is $\tau_{ij} = -\overline{v'_v v'_j}$ (in many engineering contexts the term ``Reynolds stress'' refers to this latter quantity.) The momentum equation for the mean flow, assuming that the mean transport of momentum due to molecular viscosity is negligible compared to the transport due to turbulence, is
\beq
\partial_t \bar v_i + \bar v_j \partial_j \bar v_i = - \frac{1}{\rho} \partial_i P + \partial_j \tau_{ij},
\eeq
analogous to the momentum equation, eqn.~(\ref{eq:momentum}). Also analogously, there is a turbulent flux of energy, ${\bf f}_t$, with components 
\beq
f^{(t)}_i = - \bar v_j R_{ij}
\eeq

The specific kinetic energy of the mean flow is $K = \bar v_i \bar v_i/2$. For sufficiently intense turbulence -- {i.e.} at high Reynolds number when the molecular viscosity makes a negligible contribution to the mean stress compared to the turbulence -- this obeys the transport equation
\beq
\partial_t K + \bar v_j \partial_j K = -\bar v_j \partial_j \bar P + \partial_j\left( \bar v_i \tau_{ij} \right) - S_{ij} \tau_{ji},
\eeq
which is analogous to eq.~(\ref{eq:ke}).
Here $S_{ij}$ is the symmetrized velocity gradient of the mean flow: $S_{ij} = (\partial_i {\bar v}_j + \partial_j {\bar v}_i)/2 $.

Without any modeling to simplify matters, the transport equation for the specific Reynolds stress $\tau_{ij}$ is fairly involved. Using standard symbols:
\beq
\partial_t \tau_{ij} + \bar v_k \partial_k \tau_{ij} + \tau_{ik} \partial_k \bar v_j + \tau_{jk} \partial_k v_i = \epsilon_{ij} - \Pi_{ij} + \partial_k \left[ \nu \partial_k \tau_{ij} + C_{ijk} \right]
\label{eq:hydro_R1}
\eeq
where
\beq
\Pi_{ij} = \overline{ \frac{P'}{\rho} \left( \partial_j v'_i + \partial_i v'_j \right) },
\eeq
\beq
\epsilon_{ij} = 2 \nu \overline{ (\partial_k v'_i) (\partial_k v'_j) },
\label{eq:epsilon_tensor}
\eeq
and
\beq
\rho\, C_{ijk} = \rho \overline{ v'_i v'_j v'_k } + \overline{ P' v'_i } \delta_{jk} + \overline{ P' v'_j } \delta_{ik}.
\eeq

The transport equation for the specific turbulent kinetic energy $k = \overline{v'_i v'_i}/2$ is obtained upon taking the trace, simplifying matters somewhat:
\beq
\partial_t k + \bar v_j \partial_j k = \tau_{ij} \partial_j \bar v_i - \nu \overline{ (\partial_k v'_i)(\partial_k v'_i) } + \partial_j \left[ \nu \partial_j k - \frac{1}{2} \overline{ v'_i v'_i v'_j } - \frac{1}{\rho} \overline{ P' v'_j} \right].
\eeq
Let us consider one of the simplest practical turbulence models, a one-equation model. Modeling begins when we then invoke the turbulent viscosity assumption
\beq
\tau_{ij} = 2 \nu_t S_{ij} - \frac{2}{3} k \delta_{ij}
\label{eq:blah}
\eeq
and model the sum of the turbulent (self-)transport and pressure diffusion terms (which arose from $\Pi_{ij}$ and $C_{ijk}$) as a simple diffusion term
\beq
\frac{1}{2} \overline{ v'_i v'_i v'_j } + \frac{1}{\rho} \overline{ P' v'_j } = -\frac{\nu_t}{\sigma_k} \partial_j k
\eeq
where $\sigma_k$ is a constant of order unity and indeed is often set equal to $1$ identically.

The final ingredients of the model are to assume that 
\beq
\epsilon_{jj} = 2 \nu \overline{  (\partial_k v'_j) (\partial_k v'_j) } \simeq 2 \nu \overline{ s'_{ik} s'_{ki} } = \epsilon
\eeq
where $\epsilon$ is the familiar turbulence dissipation rate (and $s_{ij}$ is the symmetrized turbulent velocity gradient), and further to assume that $\epsilon$ is related to $\nu_t$ and $k$ through
\beq
\epsilon = C_D \frac{k^2}{\nu_t}.
\label{eq:def_eps}
\eeq
Note that $\epsilon$ is a true dissipation, representing the action of molecular viscosity on small-scale eddies. In other words, $\rho \epsilon$ here is identically equal to the local heating rate ${\cal q}$.

From this we arrive at the one-equation turbulence model
\beq
\partial_t k + \bar v_j \partial_j k = \tau_{ij} \partial_j \bar v_i -  C_D \frac{k^2}{\nu_t}  + \partial_j \left(  \nu_t \partial_j k \right) = \frac{1}{\rho} \left( {\cal p} - \rho \epsilon + {\cal d} \right)
\label{eq:my_PE}
\eeq
where ${\cal p}$ is the turbulence production rate
\beq
{\cal p} = \rho S_{ij} \tau_{ji}
\label{eq:def_p}
\eeq
and ${\cal d}$ is a diffusion term. Eqn.~({\ref{eq:def_p}) should be compared with eq.~(\ref{eq:diss2}) and eq.~(\ref{eq:heat_simple}). Note that the stress tensor is symmetric in all cases so it does not matter if we symmetrize the velocity gradient in these contractions.
Eqn.~(\ref{eq:my_PE}) is a slight variation of the standard Prandtl model \citep{Pr_1945}, also proposed independently by \citet{Emm_1954}, according to \citet{Wilcox_2006}.

In steady shear, ${\cal p} = \rho \epsilon - {\cal d}$.\footnote{For homogeneous free shear flows ${\cal p} / (\rho \epsilon) \ne 1$ even though ${\cal d} = 0$, but this is because $k$ grows in time $t$ in such flows. See \citet[][sec.~5.4.5]{Pope_2000}. We could imagine a steady solution to homogeneous shear in which ${\cal p} = \rho \epsilon$, but actuality this is an artifice, since the length scale of the largest eddies will continue to grow until either boundary effects or diffusion effects become important.}
Note that in general however, whether we ignore the diffusive term ${\cal d}$ or not, there is no reason to expect the local production ${\cal p}$ to instantaneously equal the local dissipation $\rho \epsilon$, and in realistic applications they differ. In fact, if they didn't, the transport equation would be pretty useless.

Modulo diffusive effects, ${\cal p}$ and $\rho \epsilon$ equilibrate on the time scale
\beq
\tau = k/\epsilon,
\label{eq:s_def_0}
\eeq
which can be compared to the shear time scale $\gamma^{-1}$ by forming the product $\tau \gamma$.
Suppose rectilinear shear in which $v_x = \gamma y$ and all other components of ${\bf v}$ vanish. Then from eq.~(\ref{eq:def_p}) and eq.~(\ref{eq:blah}) the production rate is
\beq
{\cal p} = \rho \nu_t \gamma^2.
\eeq
Suppose as well for the moment that, suspending objections, ${\cal p} = \rho \epsilon$, then combining eq.~(\ref{eq:def_eps}) and eq.~(\ref{eq:s_def_0}), we have
\beq
\tau \gamma = C_D^{-1/2}.
\eeq
Empirical fits to laboratory turbulence suggest $0.07 \lesssim C_D \lesssim 0.09$. Then $\tau \gamma \simeq 3.5$. A similar analysis based on the $k$-$\epsilon$ model would have given a nearly identical result.

In fact, let us turn to the $k$-$\epsilon$ model now. In standard form, it consists of two coupled advection-diffusion equations, one for $k$,
\beq
D_t k = \frac{1}{\sigma_k} \partial_j \left( \nu_t \partial_j k \right) + \tau_{ij} S_{ji} - \epsilon
\label{eq:k_xport}
\eeq
and one for $\epsilon$,
\beq
D_t \epsilon = \frac{1}{\sigma_\epsilon} \partial_j \left( \nu_t \partial_j \epsilon \right) + C_1 \frac{\epsilon}{k} \tau_{ij} S_{ji} - C_2 \frac{\epsilon^2}{k},
\label{eq:epsilon_xport}
\eeq
plus the prescription
\beq
\nu_t = C_\mu \frac{k^2}{\epsilon},
\eeq
just as for the one-point model (eq.~{\ref{eq:def_eps}}). Standard values are
\beq
C_\mu = 0.09\ \ \ \sigma_k = 1\ \ \ \sigma_\epsilon = 1.3\ \ \ C_1=1.44\ \ \ C_2=1.92.
\eeq
Let us focus on the two terms on the right of eqs.~(\ref{eq:k_xport}) and (\ref{eq:epsilon_xport}). Note that $\tau_{ij} S_{ji} = {\cal p}/\rho$. Then, ignoring diffusive effects, the turbulent kinetic energy density $\rho k$ grows at the rate 
\beq
D_t(\rho k) = ({\rm diffusion}) + {\cal p} - \rho \epsilon,
\eeq
and the dissipation $\rho \epsilon$ grows at the rate
\beq
D_t(\rho \epsilon) = ({\rm diffusion}) + C_1 \frac{\cal p}{\tau} - C_2 \frac{\rho \epsilon}{\tau}.  
\eeq
Given a Heaviside step-function increase in ${\cal p}$ at some time $t_0$ then, $k$ and $\epsilon$ respond in qualitatively similar manner, both growing asymptotically to a new equilibrium at a relaxation time of order $\tau$. Again, the same results obtain with the one-equation Prandtl model as well.

In the field of laboratory studies of hydrodynamic turbulence, $\tau$ is known as the turbulence time or the turbulence dissipation time. Note also that the shear rate $\gamma$ is conventionally denoted ${\cal S}$ in turbulence studies, and $\dot \gamma$ in the theory of non-Newtonian (e.g. viscoelastic) fluids. The product $\gamma \tau$, {\it i.e.} ${\cal S}k/\epsilon$, is an important quantity in turbulence modeling; depending on the model, the value in steady shear typically lies in the range $3.5 \lesssim {\cal S}k/\epsilon \lesssim 9$. Models such as the one-equation model above and the $k$-$\epsilon$ model tend to have values of ${\cal S}k/\epsilon$ on the low end of the range. Full stress-transport models tend towards the higher end.

The relationship of the normal stress difference $(\tau_{yy} - \tau_{xx})$ to the relaxation time can not be studied in simple models such as eq.~(\ref{eq:my_PE}) or the $k$-$\epsilon$ or $k$-$\omega$ models because the normal stress difference in such models is identically zero. Let us turn now therefore to a popular Reynolds stress model of turbulence, the Launder-Reece-Rodi (LRR) model \citep{LaReRo_JFM_1975}.

The LRR model (and related similar models) model the dissipation term~(\ref{eq:epsilon_tensor}) as proportional to the simple scalar $\epsilon$ times $\delta_{ij}$, and just as for $k$-$\epsilon$, there is a transport equation for $\epsilon$. The transport term $C_{ijk}$ is modeled as a symmetrized gradient of the stress $\tau_{ij}$ and is essentially turbulent self-diffusion. The most troublesome term is $\Pi_{ij}$, the pressure-strain redistribution term. 

The LRR model for $\Pi_{ij}$ includes a so-called slow pressure strain term that follows a \citet{Rotta_1951} return-to-isotropy prescription
\beq
\Pi_{ij}^{({\rm slow})} = C_1 \frac{\epsilon}{k} \left( \tau_{ij} + \frac{2}{3} k \delta_{ij} \right).
\eeq
\citet{LaReRo_JFM_1975} give $C_1 = 1.5$. It is not so simple as to say that $\Pi_{ij}$ just dissipates or creates anisotropy however; rather, it re-orients the anisotropy, reducing the shear-aligned component but increasing other components.

The Reynolds anisotropy tensor $a_{ij}$ is here defined as
\beq
a_{ij} = \frac{-\tau_{ij} - \frac{2}{3}k \delta_{ij}}{k}.
\eeq
\citet{HamDam_PoF_2008} note an approximate evolution equation for the anisotropy is
\beq
\frac{D a_{ij}}{Dt} = - \frac{1}{\Lambda_m} a_{ij} + \alpha_2 S_{ij}
\eeq
where $\Lambda_m$ is the turbulence memory time scale,
\beq
\Lambda_m = \frac{1}{\alpha_1} \frac{k}{\epsilon}
\eeq
where 
\beq
\alpha_1 = \frac{{\cal p}}{\rho \epsilon} - 1 + C_1.
\eeq
Using $C_1 = 1.5$ and ${\cal p}/\rho/\epsilon = 1.8$, $\Lambda_m \simeq \tau / 2.3$. For LRR, $\gamma \tau = 4.83$ \citep{AbiSpe_PoFA_1993}. Then, we have
\beq
\gamma \Lambda_m \simeq 2.1.
\eeq
In other words, the isotropization rate is of order the shear rate. Qualitatively similar conclusions might be gleaned from the magnitude of the Bradshaw's (or Townsend's) constant $\beta_r = \tau_{xy} / k$; empirically, $\beta_r \simeq 0.3$, according to \citet{Wilcox_2006}, citing \citet{Tow_1976}. One can identify $s$ with $\Lambda_m$, which then justifies the statement made earlier that $s\gamma$ is of order unity (actually $2.1$ here). The resultant anisotropy leads to
\beq
\frac{\tau_{xx}-\tau_{yy}}{|\tau_{xy}|} = 1.46
\eeq
for steady shear in the LRR model, versus $2.19$ as determined experimentally \citep{AbiSpe_PoFA_1993}.

The point of this section is not to discuss these points {\em ad nauseum}, but to put some substance behind the claim made in previous work \citep{Will_NA_2004} that, in contrast to MHD turbulence, the stress anisotropy in hydrodynamic turbulence relaxes on a time scale comparable to the shear time scale, {\it i.e.} $\gamma s$ is of order unity, and that this result is tied to the result that the normalized first normal stress difference is also of order unity. These two results, in other words, are connected. Further, turbulent energy also decays on a nonzero timescale $\tau \ge s$. Ultimately, these are all empirical statements, but it is good to see how they are related to aspects of the modeling process.

\section{Appendix III: MMF Model for MHD Turbulence in Steady Fast In-spiral}
Here I discuss the purely kinematical problem of a steady solution to the MMF model for fast incompressible in-spiral. Having solved this, it is possible to discuss the resultant forces, but of course there is no guarantee that the system represents a viable solution to the full fluid equations with realistic EOS and overall force balance. Still, the example helps elucidate some important physics.

Assume an incompressible flow-field. This assumption is adopted because it allows for an exact solution that helps clarify the different roles of the normal stresses versus the off-diagonal stresses; the gross conclusions regarding the importance of the relaxation time carry over, it is expected, to the more general case of a compressible flow.
The imposed flow-field is $v_{\langle \phi \rangle} = R v^\phi = R \Omega$ where $0 < \Omega = \Omega_* (R/R_*)^{-q}$, and $v_{\langle R \rangle} = v^R = - \vartheta R_*^2 \Omega_* / R$ where $\vartheta$ is a constant, equal to the ratio $-v_{\langle R \rangle}/v_{\langle \phi \rangle}$ evaluated at $R_*$. Assume $v_{\langle z \rangle} = 0$, and that the system has azimuthal and vertical symmetry so that all quantities depend only on $R$. One can imagine the accretion flow is of some fixed vertical thickness, and within the flow, the density $\rho$ is fixed. 

The solution to the Maxwell model for the Faraday stress for the flow-field, in cylindrical coordinates, and applying an outer boundary condition that the stress asymptotes to the purely azimuthal flow solution (and therefore approaches zero at radial infinity), 
is that the off-diagonal stress $M^{R\phi}$ and the two normal stresses $M^{RR}$ and $M^{\phi \phi}$ are nonzero; the remaining stresses are zero: $M^{Rz} = M^{\phi z} = M^{zz} = 0$.
The non-zero stresses satisfy the differential equations
\begin{align}
s v^R \left( R^2 M^{RR} \right)' + R^2 M^{RR} &= - 2 \mu_t R v^R \\
s v^R \left(R M^{R\phi}\right)' - s R \Omega' M^{RR} + R M^{R\phi} &= \mu_t R \Omega' \\
s v^R R^2 (M^{\phi\phi})' - 2 s R \Omega' \left( R M^{R\phi}\right) + R^2 M^{\phi\phi} &= 2 \mu_t \frac{v^R}{R} 
\end{align}
where again, primes indicate differentiation with respect to $R$.
Note that the physical components are $M_{\langle R R \rangle} = M^{RR}$, $M_{\langle R \phi \rangle} = R M^{R \phi}$, and $M_{\langle \phi \phi \rangle} = R^2 M^{\phi \phi}$.
Now assume that the relaxation rate $s^{-1}$ is proportional to the shear rate $q\Omega$, and define the constant $\hat s = s \Omega$. Also write
\beq
x \equiv \frac{R}{R_*}\ \ \ w_q(x;\vartheta,\hat s) \equiv \frac{M^{RR}}{\mu_t \Omega_*}\ \ \ y_q(x;\vartheta, \hat s) \equiv \frac{-RM^{R\phi}}{\mu_t \Omega_*}\ \ \ z_q(x;\vartheta,\hat s) \equiv \frac{R^2 M^{\phi \phi}}{\mu_t \Omega_*}.
\eeq
Up to sign convention, the functions $w_q$, $y_q$ and $z_q$ are the physical stresses scaled to the fiducial value $\mu_t \Omega_*$.
Now set $q=1$. Actually, it turns out that $w$ is independent of $q$, once we zero out the constant of integration, {\it i.e.} toss out the unphysical homogeneous solution, which is ruled out by the imposed boundary conditions. In any case,
\beq
w_1(x;\vartheta) = 2 \vartheta x^{-2}.
\eeq
Define $\tilde x \equiv x / (\hat s \vartheta)$. The remaining solutions may be found by using integrating factors to turn the differential equations into exact differentials. The solution for $y_1$ is that
\beq
y_1(x;\vartheta,\hat s) = \frac{1}{x} \left[ 2 - \tilde x e^{\tilde x} E_1(\tilde x) \right]
\eeq
where $E_1$ is the exponential integral,
\beq
E_1(x) = \int_x^\infty \frac{e^{-t}}{t}\ dt.
\eeq
The solution for $z_1$ is
\beq
z_1(x;\vartheta,\hat s) = 2 \vartheta^{-1} {\tilde x}^2 e^{\tilde x} \left[ - \frac{1}{\tilde x} E_1(\tilde x) + \Gamma(-1,\tilde x) + 2 \Gamma(-2, \tilde x) - \frac{1}{{\hat s}^2} \Gamma(-3,\tilde x) \right],
\eeq
where $\Gamma(a,x)$ is the upper incomplete Gamma function,
\beq
\Gamma(a,x) = \int_x^\infty e^{-t} t^{a-1}\ dt.
\eeq
Again, in the cases of $y_1$ and $z_1$ just as for $w_1$, there is also a homogeneous solution that comes from the constant of integration that arises upon integrating the exact differential obtained by multiplying the differential equations by the integrating factor. These homogeneous solutions can again be tossed out, using the imposed boundary conditions at large $R$.

Note that for $R \gg R_*$ and for small $\hat s$ and vanishing $\vartheta$, the solution above approaches the purely azimuthal flow-field ($v_{\langle R \rangle} = 0$) solution
\begin{align}
M^{RR} = M_{\langle R R \rangle} &= 0, \\
R M^{R \phi} = M_{\langle R \phi \rangle} &= \mu_t R \Omega', \\
R^2 M^{\phi \phi} = M_{\langle \phi \phi \rangle} &= 2 \mu_t s R^2 \left( \Omega' \right)^2.
\end{align}
Because of the particular flow-field chosen in which $v^z=0$, upper and lower streamsurfaces of this flow are parallel to the $z=0$ plane, plus, vertical integration to find ${\cal R}$, ${\cal P}$ and ${\cal W}$ is straightforward. These quantities are discussed and shown in the main text.

\bibliographystyle{aasjournal}
\bibliography{no_bdy_layer}{}

\newpage

\end{document}